\documentclass[a4paper,12pt]{article}

\usepackage[scale={.8,.8}]{geometry}
\usepackage[intlimits]{amsmath}
\usepackage{amssymb}
\usepackage{enumerate}
\usepackage{cite} 
\usepackage{a4wide}

\usepackage{euscript}
\usepackage{amsfonts}
\usepackage{amsbsy}
\usepackage{epsfig}

\newcommand{\Zbb}{\mathbb{Z}}

\newcommand{\Pbb}{\mathbb{P}}

\numberwithin{equation}{section}
\numberwithin{table}{section}
\begin{document}

\title{ 
{\normalsize \vspace*{-1.5cm} DESY 12-130     
\hfill\mbox{}\\
ZMP-HH/12-15\hfill\mbox{}\\
\vspace*{0.1cm} August 2012\hfill\mbox{}\\}
\vspace{1cm}
\bf 
{\huge Voisin-Borcea Manifolds \vspace{4mm} \\ and\\   \vspace{4mm} Heterotic Orbifold Models}
\vspace{1cm}
\author{W.~Buchmuller$^a$ , J.~Louis$^{b,c}$, J.~Schmidt$^a$, R.~Valandro$^b$
\\[2mm]
{\normalsize\itshape  $^a$~Deutsches Elektronen-Synchrotron~DESY,} \\ 
{\normalsize\itshape Hamburg, Germany}\\
{\normalsize\itshape  $^b$~II Institute for Theoretical Physics,}\\
{\normalsize\itshape Hamburg University, Hamburg, Germany}\\
{\normalsize\itshape  $^c$~Zentrum f\"ur Mathematische Physik,}\\
{\normalsize\itshape Hamburg University, Hamburg, Germany}\\[2mm]
{\normalsize {\ttfamily buchmuwi@mail.desy.de}, {\ttfamily jan.louis@desy.de},}\\{\normalsize{\ttfamily schmijon@mail.desy.de}, {\ttfamily roberto.valandro@desy.de}}}}
\date{}
\maketitle

\begin{abstract}
\noindent
We study the relation between a heterotic $T^6/\Zbb_6$ orbifold  model and a compactification on a smooth Voisin-Borcea Calabi-Yau three-fold with non-trivial line bundles. This orbifold can be seen as a $\Zbb_2$ quotient of $T^4/\Zbb_3\times T^2$.
We consider a  two-step resolution, whose intermediate step is $(K3\times T^2)/\Zbb_2$. This allows us to identify the massless twisted states which correspond
to the geometric K\"ahler and complex structure moduli. We work out the match 
of the two models when non-zero expectation values are given to all twisted
geometric moduli. We find that even though the orbifold gauge group contains an
$SO(10)$ factor, a possible GUT group, the subgroup after higgsing does
not even include the standard model gauge group.
Moreover, after higgsing, the massless spectrum is non-chiral under the 
surviving gauge group.

\end{abstract}

\vfill

\thispagestyle{empty}

\newpage
\tableofcontents

\newpage

\section{Introduction}

An important corner of the space of string vacua is populated by string orbifold compactifications \cite{Dixon:1985jw,Dixon:1986jc,Katsuki:1989bf}.
In recent years, a landscape of phenomenologically interesting
heterotic orbifold models has been constructed, which contain gauge group and
chiral matter of the supersymmetric standard model. A large class of these models is based on 
the $E_8\times E_8$ heterotic string compactified on the
$T^6/\Zbb_{6-II}$ orbifold
\cite{Kobayashi:2004ya,Buchmuller:2005jr,Buchmuller:2006ik,Lebedev:2006kn,Lebedev:2008un,Blaszczyk:2009in}. 
Employing worldsheet conformal field theory techniques
the spectrum and some couplings in the effective low energy action
have been computed.

On the other hand, these models also have `unwanted' features
such as a large number of massless states which form real representations
with respect to the standard model gauge group but are chiral with
respect to the full orbifold gauge group. To obtain realistic models
one has to move away from the orbifold by higgsing 
part of the gauge group, which also generates masses for the vector-like 
part of the spectrum. In practice, this is rather cumbersome and the
obtained results suffer from the fact that the exact form of the 
superpotential is unknown.

The appearance of vector-like massless fields might be an artifact of the 
construction at the orbifold
point where generically additional symmetries and additional massless
states appear. Therefore it is of interest to move away from the
orbifold locus and construct the theories at generic points in their
moduli space. Geometrically, this corresponds to blowing up the
orbifold singularities and identifying smooth Calabi-Yau compactification 
manifolds. In this way, one may hope to relate the massless vector-like fields
to geometric moduli and obtain a better understanding of their mass generation
and the stabilization of the geometry. 

The relation between heterotic orbifold models and
their smooth counterparts has been actively investigated in recent 
years \cite{ht06,Nibbelink:2007rd,Nibbelink:2007pn,nkpt08,nt09,bn10,Blaszczyk:2011ig}. 
First it was studied from a local point of
view: The orbifold singularity was blown up using toric methods
\cite{lr06} and then the match of the spectrum with the corresponding
smooth compactification was established. 
Subsequently this was generalized to globally defined, compact 
orbifold backgrounds by
gluing together patches related to
different singularities to a Calabi-Yau three-fold with a gauge bundle
\cite{nt09}. 
Starting from standard model-like orbifold models on
$T^6/\Zbb_{6-II}$, it was found that the hypercharge $U(1)$ gauge boson
always becomes massive in the blow up process. In \cite{bn10}, this problem 
could be avoided by considering the orbifold $T^6/(\Zbb_2)^3$ where one of the $\Zbb_2$ factor acts freely on $T^6$.
Recently, also Gauged Linear Sigma Models have been used
to construct resolutions of toroidal orbifolds \cite{NibbelinkGroot:2010wm,Blaszczyk:2011ib,Blaszczyk:2011hs}. 

In this paper we study a specific $T^6/\Zbb_{6-II}$ orbifold model
which was previously analyzed from the perspective of grand unification in
six dimensions \cite{Buchmuller:2007qf,Buchmuller:2008uq}. 
The orbifold under consideration can be viewed as 
the space $(T^4/\Zbb_3\times T^2)/\Zbb_2$, with an intermediate
six-dimensional theory corresponding to the compactification on
$T^4/\Zbb_3$. For an anisotropic compactification, with the size of $T^4$
small compared to the size of $T^2$, the latter can be related to the
scale of grand unification.  The four-dimensional theory is then obtained by 
further compactifying on $T^2/\Zbb_2$. For an example with non-zero Wilson
lines it was shown that this two step procedure leads to the same
spectrum as the direct 
compactification on $T^6/\Zbb_{6-II}$ \cite{Buchmuller:2007qf}.
To simplify the analysis we
turn off all Wilson lines in the present paper.

Our goal is to
identify a smooth manifold corresponding  to the $T^6/\Zbb_{6-II}$
orbifold and to determine its properties. In order to do so we again employ 
 a two-step analysis. First we resolve $T^4/\Zbb_3$ to a
smooth K3 surface and use the fact that the match of the spectra
has already been established in \cite{ht06}. In a second step we
compactify on $T^2$ and divide out the $\Zbb_2$ action. This gives 
the singular space $(K3\times T^2)/\Zbb_2$ which can be resolved to 
obtain a smooth Calabi-Yau three-fold termed Voisin-Borcea manifold \cite{Voisin,Borcea}.

In the orbifold under consideration the original $E_8\times E_8$ gauge group is
broken by a specific gauge twist. In the smooth compactification
this corresponds to the presence of a non-trivial gauge bundle. 
From the structure of the gauge bundle one can  infer 
which twisted states correspond to geometric moduli and furthermore 
see that the transition from the
orbifold to the smooth manifold is equivalent to a motion in the field
space of twisted states. 
If these states 
are charged with respect to the gauge group at the orbifold point,
spontaneous symmetry breaking occurs leaving a
smaller gauge group in the smooth compactification. This allows us
to identify which {\it massless} twisted states correspond to
 geometric K\"ahler moduli. This procedure differs from \cite{nt09},
where some K\"ahler moduli were identified with massive twisted
states.
Using properties of the Voisin-Borcea manifold 
we are also able to determine which 
twisted states of the orbifold correspond to geometric 
complex structure moduli. 

We find that the vacuum expectation values (VEVs) 
needed to blow up the orbifold completely, 
break the orbifold gauge group to a small
subgroup. In fact, to go to a smooth point in moduli space, all 
massless twisted states which correspond to geometric moduli have to acquire 
non-zero VEVs. Giving VEVs to all of these states breaks the orbifold gauge group, reducing consistently its rank. It turns out that
the resulting light spectrum is {\it non-chiral} with respect to the unbroken gauge group. 
This perfectly matches with the results obtained for smooth compactifications.

The paper is organized as follows. In Section \ref{Sec6D} 
we review the correspondence between  
compactifications on the $T^4/\Zbb_3$ orbifold and a specific $K3$ background with a line bundle. In this example we 
describe the techniques 
involved in matching the two backgrounds.
In Section \ref{Sec4D} we turn to 
the four-dimensional backgrounds and use similar techniques 
to show the correspondence between orbifold and smooth compactifications.
 We first describe how to resolve the singular space $Y_s=(K3\times T^2)/\Zbb_2$ to a smooth Voisin-Borcea Calabi-Yau three-fold. After the identification of the proper gauge bundle, we give the map between the geometric moduli and the corresponding twisted states. We show that the resulting gauge group is the same and that the non-Abelian massless spectrum also matches.
The three appendices give further details on K3 surfaces and blow-up of
singularities, and on the massless spectra in four dimensions, for orbifold
and Calabi-Yau three-fold.

\section{Heterotic Compactifications to $D=6$}\label{Sec6D}

In this section we  consider the $E_8\times E_8$ heterotic string
compactified on a four dimensional internal space. Specifically, 
we first study the compactification on the orbifold $T^4/\Zbb_3$
with a particular gauge twist but
without Wilson lines. Subsequently we analyze the compactification of 10D
supergravity on the resolved K3 surface with an
appropriate gauge background. These backgrounds and the corresponding
matching was studied in \cite{ht06} and here we briefly discuss their results.

\subsection{Heterotic compactification on $T^4/\Zbb_3$}\label{HetT4}

In absence of Wilson lines, a heterotic orbifold model is completely specified by the action of the orbifold group on the internal geometry and on the gauge bundle. Let us start with the geometry.

\subsubsection*{Geometry of $T^4/\Zbb_3$}

The action of a $\Zbb_N$ orbifold group on a four dimensional torus $T^4$ is specified by the twist vector $v=(v_1,v_2)$, where $N\cdot v_i \in \Zbb$. 
The $v_{i=1,2}$ determine the transformation of the two complex
coordinates $z_i=x_i+\tau_i y_i$ 
of $T^4$: 
\begin{equation}\label{deftheta}
\theta:\quad  z_i \mapsto e^{2\pi i v_i} z_i \ ,
\end{equation}
where $\tau_{i}$ denote the two complex structure parameters of $T^4$.
To have a supersymmetric background the $v_i$ must satisfy $v_1+v_2=0$ mod $1$. 
In the following we consider the specific case $N=3$ and
$v = \left( -\tfrac13, \tfrac13 \right).$
In order to have a $\Zbb_3$ symmetric $T^4$, $\tau_1$ and $\tau_2$ are fixed to generate the Lie algebra lattice of $G_2\times SU(3)$.

\begin{figure}[b]
\begin{center}
\includegraphics[width=8cm]{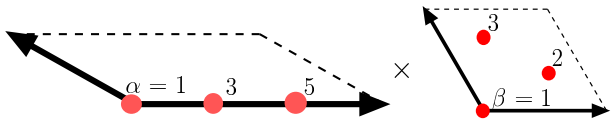}\\
\end{center}
\caption{\it $\Zbb_3$ fixed points on $T^4=T^2\times T^2$. The 2-torus on the left is labelled by $z_1$, while the 2-torus on the right is labelled by $z_2$.}\label{T4Z3fig1}
\end{figure}

The resulting geometry is a singular space with nine $\mathbb{C}^2/\Zbb_3$ orbifold singularities (see Fig.~\ref{T4Z3fig1}) at
\begin{equation}\label{singularpointsT4}
  (z_1,z_2) = (z_1^{\alpha},z_2^{\beta})\qquad\qquad \mbox{with } \,\,\, \alpha=1,3,5 \,\,\, \mbox{ and }\,\,\, \beta=1,2,3 \:. 
\end{equation}
These are the points of $T^4$ left fixed by the $\Zbb_3$ action.

Furthermore, there are four two-cycles of $T^4$ surviving the orbifold
projection. Expressed in terms of their Poicar\'e dual
two-forms they are given by
\begin{equation}\begin{aligned}\label{four2}
  \Pi_1 &= P\!D[3dx_2\wedge dy_2] \ , \qquad 
 \Theta_1 = P\!D[-dx_1\wedge (2dx_2+dy_2)-3dy_1\wedge dx_2] \ ,\\
  \Pi_2 & = P\!D[3dx_1\wedge dy_1] \ , \qquad \ \
 \Theta_2 = P\!D[ dx_1 \wedge (dx_2-dy_2) -3 dy_1\wedge dy_2] \ ,
\end{aligned}\end{equation}
where $P\!D[\cdot]$ indicates the  Poincar\'e dual of the two-form in the bracket.
$\Pi_1$ and $\Pi_2$ are the two $T^2$ spanned by $z_1$ and $z_2$. 
The other two two-cycles are 
combinations of the integral two-forms constructed from
$dx_1,dy_1$ of the first $T^2$  and $dx_2,dy_2$ of the second $T^2$.
One can check that all four two-cycles
are invariant under $\Zbb_3$ which acts on the 1-forms of $T^4$ as
\begin{equation}
 dx_1\mapsto -2dx_1-3dy_1 \ , \qquad dy_1\mapsto dx_1+dy_1  \ ,
\qquad dx_2\mapsto dy_2 \ , \qquad dy_2\mapsto -dx_2-dy_2 \ .
\end{equation}
The four two-cycles  
given in \eqref{four2} have the intersection matrix:
\begin{equation}\label{twocyc}
  \left( \begin{array}{ccrr}  0 & 3 && \\ 3 & 0 && \\  && 2 & -1 \\ &&
        -1 & 2 \\ \end{array}\right)\ ,
\end{equation}
where the first  block contains $\Pi_1, \Pi_2$ while the second
contains $\Theta_1,\Theta_2$. Note that the normalization of (\ref{twocyc})
is set by $\int_{T^4} dx_1dy_1dx_2dy_2 = 1$.
Finally, the K\"ahler form $j$ 
and the holomorphic two-form $\Omega_2$ are given by
\begin{equation}\label{jomegadef}
 j = t_1 dx_1\wedge dy_1 + t_2 dx_2\wedge dy_2\ , \qquad \qquad \Omega_2
 = dz_1\wedge dz_2 \ ,
\end{equation}
where $t_{1,2}$ are two K\"ahler parameters of $T^4$.

\subsubsection*{Gauge twist and massless spectrum}

The $\Zbb_N$ orbifold group also acts on the $E_8\times E_8$ gauge
degrees of freedom by means of the twist vector
\begin{equation}\label{Vdef}
 V = \tfrac{1}{N}\left(n_1,...,n_8)(n_9,...,n_{16} \right) \ ,  
\end{equation}
where the first (second) bracket acts on the first (second) $E_8$
factor 
and all $n_{a=1,\ldots,16}$ are integers or half-integers.
States with weight vector $w$ transform under the orbifold action 
with the phase $e^{2\pi i V\cdot w}$.
Modular invariance of the worldsheet partition function requires 
a relation between $V$ and the vector $v$ defined in the previous
section
\begin{equation}
 N \left(\sum_{a=1}^{16} V_a^2 - \sum_{i=1}^2 v_i^2 \right) = 0 \mbox{ mod } 2\ .
\end{equation}
For the $T^4/\Zbb_3$ orbifold at hand, one can 
choose the gauge twist
\begin{equation}\label{T4twist}
 V = (0^2, \tfrac23,0^5)(\tfrac23,0^7 ) \ ,
\end{equation}
which breaks the gauge group from $E_8\times E_8$ to
\begin{equation}\label{6DorbGG}
  G = SO(14)\times U(1) \times SO(14)\times U(1)  \ .
\end{equation}
The two $U(1)$'s have the following generators:
\begin{equation}
 T_1 = (0^2,2,0^5)(0^8)\ , \qquad\quad T_2 = (0^8)(2,0^7)\ .
\end{equation}

The massless spectrum can be organized in 6D ${\cal N}=1$
supermultiplets. 
In the untwisted sector of the orbifold one finds hypermultiplets
in the gauge representations
\begin{equation}
 ({\bf 64},{\bf 1})_{1,0} \,+\, ({\bf 14},{\bf 1})_{-2,0}  \,+\, ({\bf
   1},{\bf 64})_{0,1}  \,+\, ({\bf 1},{\bf 14})_{0,-2} \,+\, 2({\bf
   1},{\bf 1})_{0,0} \ ,
\end{equation}
where the first (second) entry corresponds to the first (second) 
$SO(14)$ factor and the subscripts denote the two $U(1)$ charges.
In addition one obtains the usual supergravity and dilaton
multiplets, including the antisymmetric tensor $B$, and two hypermultiplets
containing the `radion' fields of the two tori.  
The twisted states are  
localized at the nine fixed points
$(z_1^{\alpha},z_2^{\beta})$. At each fixed point there is a
twisted spectrum with representations
\begin{equation}\label{spect}
 ({\bf 14},{\bf 1})_{-\tfrac23,\tfrac43}  \,+\, ({\bf 1},{\bf
   14})_{\tfrac43,-\tfrac23} \,+\, 2 ({\bf 1},{\bf
   1})_{\tfrac43,\tfrac43} \ .
\end{equation}

The 6D low energy theory around this orbifold vacuum contains a D-term
potential for the scalars in the hypermultiplets. 
For canonically normalized scalars it reads
\begin{equation}
 V = \frac{1}{2}\sum_{i,A} D^{i,A} D^{i,A} \ , \qquad\qquad \mbox{where} \qquad  D^{i,A} = \Phi_m^\dagger \sigma^i T^A_{mn} \Phi_n \ .
\end{equation}
Here the $\Phi_n$ are doublets of complex scalars $(\phi_n,\tilde{\phi}_n)$,
$\sigma^i$ are the Pauli matrices
and $T^A$ the generators of the gauge group. If the gauge group
contains $U(1)$ factors, then the corresponding generators are
$T^A_{mn}=q_n^A \delta_{mn}$. 

As we will see shortly,
the correspondence with the smooth $K3$
compactification
can be established by giving a VEV to a particular flat direction of the potential. Anticipating the result let us consider  
the singlets of $SO(14)\times
SO(14)$ whose potential is solely determined by the $D$-terms of
$U(1)\times U(1)$ . From \eqref{spect} we see that
at each singularity $(z_1^{\alpha},z_2^{\beta})$
there are two states $\Phi_0^{\alpha\beta}$ and $\Phi_1^{\alpha\beta}$ in the representation
$({\bf 1,1})_{4/3,4/3}$ which are charged under both $U(1)$s. 
This implies that flat directions of the potential are 
\begin{equation}\label{6dfirstFD}
 \langle(\phi_0^{\alpha\beta},\tilde{\phi}_0^{\alpha\beta})\rangle =
 (v^{\alpha\beta},0) \ ,\qquad\quad
	\langle(\phi_1^{\alpha\beta},\tilde{\phi}_1^{\alpha\beta})\rangle = (0,v^{\alpha\beta}) \ ,
\end{equation}
which break $U(1)\times U(1)$ to the diagonal $U(1)$,
leaving $ SO(14)\times SO(14)\times U(1) $
as the unbroken gauge group. The vector multiplet
of the broken $U(1)$ eats one
hypermultiplet in the singlet representation of the unbroken group
to give a massive vector multiplet. Note that there are also other flat
directions which break  $ SO(14)\times SO(14)\times U(1) $. 

In order to facilitate the comparison with the smooth compactification
in the next section, it is convenient to define new field
variables $\varphi^{\alpha\beta}_i$ by the transformation
\begin{equation}\label{redefT}
 \phi_0^{\alpha\beta} \equiv v^{\alpha\beta} e^{\varphi^{\alpha\beta}_0} \
 ,\qquad\qquad \tilde{\phi}_1^{\alpha\beta} \equiv v^{\alpha\beta}
 e^{\varphi^{\alpha\beta}_1}\ .
\end{equation}
If the  $\phi_0^{\alpha\beta}$ have charge $q$ and 
transform linearly under the $U(1)$ gauge transformation  
$\phi_0^{\alpha\beta}\mapsto e^{iq\chi}\phi_0^{\alpha\beta}$, 
then the field $\varphi_0^{\alpha\beta}$ transforms non-linearly with a local shift
\begin{equation}
 \varphi_0^{\alpha\beta}\mapsto \varphi_0^{\alpha\beta} + i \,q\, \chi \ .
\end{equation}
This gauged shift symmetry gives mass to the corresponding $U(1)$ vector.
The ``axionic'' fields $\varphi^{\alpha\beta}_i$ can also be used to redefine 
the other fields in the associated $(\alpha,\beta)$-twisted sector. 
This redefinition changes their charge with respect to the broken $U(1)$.
For example, multiplying the hypermultiplet $\Phi_{(1,14)}^{\alpha\beta}$
with charges (4/3,-2/3) by
$e^{-\varphi_1^{\alpha\beta}}$,\label{redefinPage} one obtains 
${\bf 14}$-plets with charges (0,-2). After analogous redefinition and including the untwisted states, one obtains the spectrum:
\begin{equation}\label{spectrumredef}
 ({\bf 64},{\bf 1})_{1,0} \,+10 \, ({\bf 14},{\bf 1})_{-2,0}  \,+\, ({\bf 1},{\bf 64})_{0,1}  \,+10\, ({\bf 1},{\bf 14})_{0,-2} \,+\, 20({\bf 1},{\bf 1})_{0,0} \:.
\end{equation}

Below the scale of symmetry breaking the unbroken gauge group only contains the diagonal $U(1)$ with charges 
\begin{equation}\label{ReducesSpHetK3}
  ({\bf 64,1})_{1} + ({\bf 1,64})_{-1} + 10 \, ({\bf
   14,1})_{-2} + 10 \, ({\bf 1,14})_{+2} + 19 \, ({\bf 1,1})_{0}  \ ,
\end{equation}
where furthermore one singlet was removed by the Higgs mechanism.

\

In addition to the flat direction considered so far there are other flat
directions which break the orbifold gauge group to smaller subgroups. The
following example is useful in order to understand the matching between 
orbifold compactifications with K3 compactifications. The vacuum 
is now characterized by expectation values
of the non-Abelian singlet fields $\Phi_0^{\alpha\beta}$ and
$\Phi_1^{\alpha\beta}$ as well as $\Phi_{(1,14)}^{\alpha\beta}$: 
\begin{equation}\label{6dsecondFD}\begin{array}{rclrcl}
   \langle(\phi_0^{\alpha \beta},\tilde{\phi}_0^{\alpha\beta})\rangle\big|_{\beta=1} &=&( \sqrt{2} v, 0)\ , \qquad &
  \langle(\phi_1^{\alpha \beta},\tilde{\phi}_1^{\alpha\beta})\rangle\big|_{\beta=2,3} &=& ( 0 , v)\ ,\\  
    \langle(\phi_{(1,14)}^{\alpha \beta},\tilde{\phi}_{(1,14)}^{\alpha \beta})\rangle\big|_{\beta=1} &=& (\sqrt{2}\hat{v}, 0)\ ,   \qquad & 
      \langle(\phi_{(1,14)}^{\alpha \beta},\tilde{\phi}_{(1,14)}^{\alpha \beta})\rangle\big|_{\beta=2,3} &=& (0, \hat{v})\ ;
\end{array}
\end{equation}
any component of the ${\bf 14}$-plet may be chosen to have non-zero VEV.
As we shall discuss in more detail in Section~3.1, each component is
characterized by a vector which is the sum of the shift vector $V$ and
a root vector, e.g.\ $\tilde{p} = V + (0^8)(-1,1,0^6)$.
One easily verifies that this configuation represents a flat direction.
One of the $SO(14)$ factors is now broken to $SO(12)$, and the surviving gauge 
group reads
\begin{equation}
 SO(14)\times SO(12)\times U(1)\ .
\end{equation}
In the above example the same component of the ${\bf 14}$-plet acquires a VEV
at all fixed points. Choosing VEVs for different components on different fixed 
points one can achieve breaking to smaller subgroups.

The VEVs given in Eqs.~\eqref{6dfirstFD} and \eqref{6dsecondFD} correspond to directions in field
space which lead away from the orbifold point into a smooth
compactification. 
As we will see in the next section, the corresponding twisted
states can be identified with the moduli controlling the sizes 
of collapsed cycles in the singularity.
 One expects that by giving large enough VEVs this procedure
can be continued into a regime where 
the supergravity approximation of the heterotic string can be trusted.

\subsection{Heterotic compactification on K3 with line bundles}\label{K3}

In this section we briefly recall compactifications of 10D supergravity
 on smooth K3 manifolds in presence of a non-trivial
gauge bundle \cite{ht06,Honecker:2006dt,Nibbelink:2008qf,Louis:2011hp}.
The goal is to find 
a set of backgrounds with appropriate flat direction that 
contain the orbifold  $T^4/\Zbb_3$ as a limit in the K3 moduli space. 

\subsubsection*{K3 moduli space}

The metric of K3 is defined by the overall
volume $\nu$ together with a three-dimensional subspace $\Sigma$ of 
the second cohomology group $H^2(K3)$ (see \cite{a96} for a review). $H^2\!\left(K3\right)$ is a 22-dimensional vector space equipped with
a natural scalar product of signature  $(3,19)$ defined as
\begin{align}\label{K3modmetric}
  (v \cdot w) \equiv \int_{K3} v \wedge w \ , \quad \qquad \forall\;
  v,w\in H^2\!\left(K3\right)\ .
\end{align}
This product must be positive definite when restricted to $\Sigma$.

Taking three orthonormal vectors $\omega_1$, $\omega_2$, $\omega_3$  spanning the subspace $\Sigma$, one can define
the K\"ahler form $j$ and the holomorphic two-form 
$\Omega_2$:\footnote{This definition is not unique: there is an $S^2$
  of possible 
complex structures and associated K\"ahler forms with each of them defining the
same metric.}
\begin{equation}\label{Sigma}
  j= \sqrt{2 \nu}\,\omega_3\ , \qquad \Omega_2=\omega_1 +i \omega_2\,.
\end{equation}

The motion in moduli space can be parametrized by the motion of the positive norm
three-plane $\Sigma$ inside $H^2(K3)$, i.e., by
$$
{\cal M}_\Sigma=\frac{SO(3,19)}{SO(3)\times SO(19)}\ .
$$
${\cal M}_\Sigma$ has 57 moduli, that together with the volume give the 58 geometric moduli of K3. Adding the 22 B-field moduli (i.e., the periods of the two-forms $B$ along the 22 harmonic two-form of K3), one obtains 80 moduli. These belong to 20 hypermultiplets of the 6D ${\cal N}=1$ supergravity coming from compactifying the heterotic theory on K3.

The position of $\Sigma$ in $H^2(K3,\Zbb)$ determines what cycles have
zero size: If a two-cycle is orthogonal to $\Sigma$, it is shrunk to zero. 
If a set of two-cycles with self-intersection $-2$ (i.e., two-spheres)
shrink, an orbifold singularity appears. These singularities are
classified by ADE-groups that are subgroups of $E_8\times E_8$. K3
develops an ADE-singularity of type $G$ if the shrinking two-spheres
have an intersection 
matrix equal to minus the Cartan matrix of the ADE group $G$.

\begin{figure}[t]
\begin{center}
\includegraphics[width=14cm]{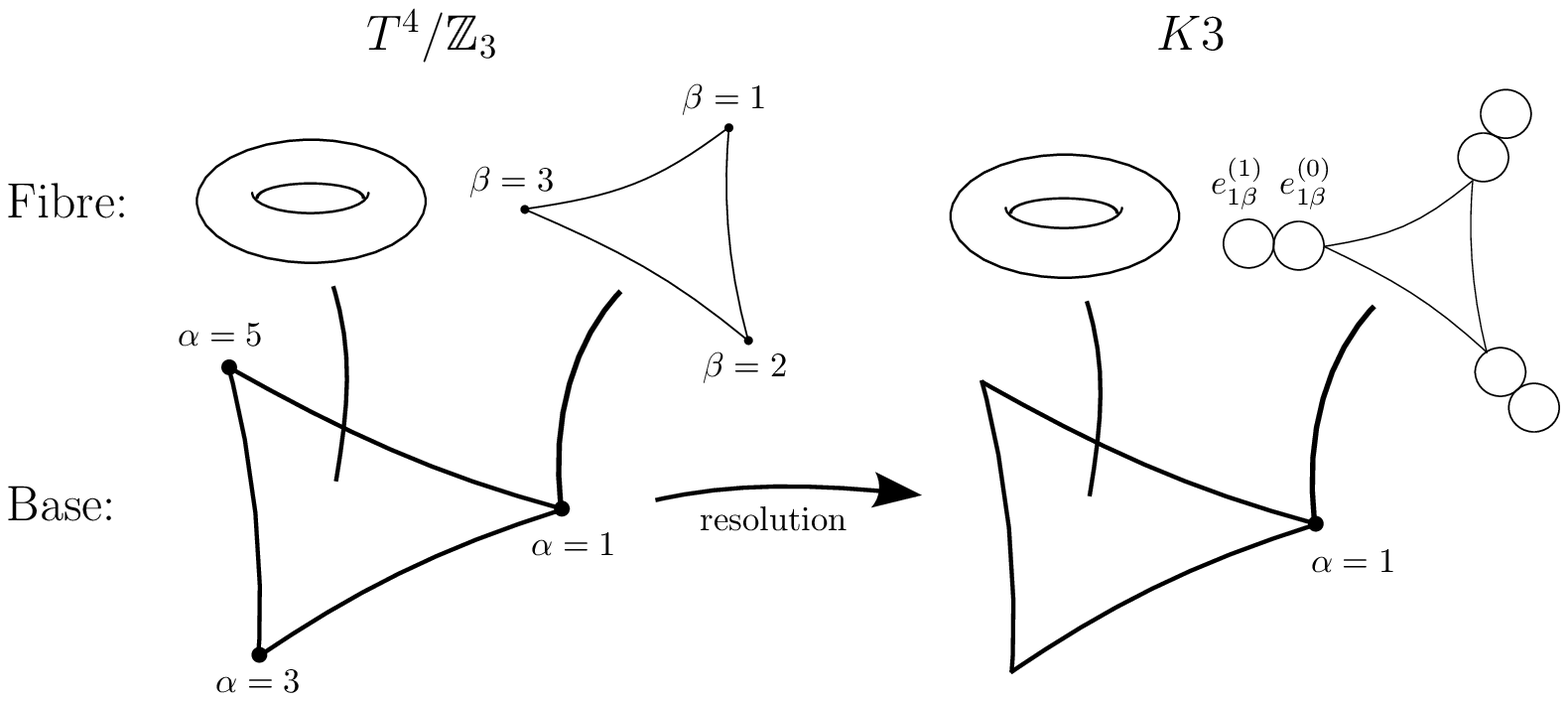}\\
\end{center}
\caption{\it From $T^4/\Zbb_3$ to a smooth $K3$}\label{K3Resol}
\end{figure}

\subsubsection*{From $T^4/\mathbb{Z}_3$ to a smooth K3}

$T^4/\Zbb_3$ is a singular K3 that is elliptically fibered over the base $T^2/\Zbb_3$ (see Fig. \ref{K3Resol}). Over the singular points $z_1=z_1^\alpha$ of the base (spanned by $z_1$), the fiber (spanned by $z_2$) degenerates to $T^2/\Zbb_3$. 
$T^4/\mathbb{Z}_3$ has nine $A_2$ singularities (i.e., $\mathbb{C}^2/\Zbb_3$ singularities). This corresponds to the limit of a smooth K3 in which a set of cycles have shrunk to zero size. More precisely, 
each $A_2$ singularity is generated by two shrinking
two-cycles $e_{\alpha\beta}^{(0)}$ and $e_{\alpha\beta}^{(1)}$ 
where $\alpha\beta$ specifies the fixed point.
Their 
intersection matrix is given by\footnote{In the following we identify two-forms and two-cycles of K3, going from one to the other by
Poincar\'e duality. On the space of two-cycles (two-forms) there is a
natural inner product given by the intersection form or \eqref{K3modmetric} respectively.} 
\begin{equation}\label{A2intesMat}
\left( \begin{array}{cc}
  -2 & 1 \\ 1 & -2 \\
 \end{array}\right)\ ,
\end{equation}
with all intersections between cycles at different singularities vanishing.

There are two two-cycles at each of the nine singularities
resulting in a total of 18 shrinking two-cycles, all of them orthogonal to
the 3-plane $\Sigma$. Since  $H^2(K3)$ is 22-dimensional,
$\Sigma$ still has the freedom to move
in a four dimensional space with signature $(3,1)$. Thus the number of geometrical moduli left is four which correspond to 
the overall volume and the parameters controlling the residual motion of $\Sigma$.

At the $T^4/\Zbb_3$ point of the moduli space, the lattice of cycles orthogonal to the holomorphic two-form $\Omega_2$ (the Picard lattice $P_X$) is (see \cite{dm96})
\begin{equation}\label{PXlattice}
 P_X = {\cal U} \oplus (-E_6) \oplus (-E_6) \oplus (-E_6) \qquad\qquad\mbox{with}\qquad
	{\cal U} = \left( \begin{array}{cc}  0 & 1 \\ 1 & 0 \\ \end{array}\right)\ 
\end{equation}
and $(-E_6)$ denoting the Cartan matrix of $E_6$.\footnote{We see the appearance of the Cartan matrix of $E_6$ in Fig.~\ref{K3Resol}: The $E_6$ cycles are the $\mathbb{P}^1$ located at, for example, $\alpha=1$ on the right side of the figure. We see that their intersection matrix is actually the same as the Cartan matrix of $E_6$.}
Therefore $P_X$ is a space of signature $(1,19)$. 
If $j$ stayed along the ${\cal U}$ part of the lattice, then the shrunken cycles would generate three $E_6$ singularities (remember that a cycle is shrunk to zero size if it is orthogonal to both $j$ and $\Omega_2$). To have nine $A_2$ singularities instead, $j$ must be orthogonal to the lattice\footnote{In order to have a $A_2$ singularity, the corresponding lattice must be primitively embedded in $P_X$ (a lattice embedding $\ell \hookrightarrow L$ is primitive if $L/\ell$ is torsion-free).}
\begin{equation}
 \Lambda^{\rm shr} = (-A_2)^{\oplus 9} \:.
\end{equation}
In fact, $j$ is a combination of the two tori $\Pi_1$ and $\Pi_2$, with 
intersection matrix
\begin{equation}
 \left( \begin{array}{cc}  0 & 3 \\ 3 & 0 \\ \end{array}\right)\ ,
\end{equation}
that can be shown to be orthogonal to the $A_2$ cycles. They are linear combinations of the
${\cal U}$-block and the $E_6$ cycles.

So far we have listed only 20 independent two-cycles of the 22 cycles of K3: $\Pi_1$, $\Pi_2$, and the 18 $A_2$ cycles. These span a sublattice of $P_X$ with the same rank. The orthogonal complement in $H_2(K3,\Zbb)$ is generated by the two cycles that span the $\Omega_2$-plane, i.e., the two cycles $\Theta_1$ and $\Theta_2$ of $T^4$ left invariant by $\Zbb_3$.

To go away from the orbifold point, $\Sigma$ has to move inside the space spanned by the cycles $e_{\alpha\beta}^{(0)},e_{\alpha\beta}^{(1)}$, in such a way that they are no more orthogonal to it. In terms of
the K\"ahler form and the holomorphic two-form, this means that they have non-zero components along the space spanned by $e_{\alpha\beta}^{(0)},e_{\alpha\beta}^{(1)}$.

\subsubsection*{Fluxes on K3}

In supergravity compactifications an additional constraint arises from the Bianchi identity of the $B$-field, which is given by\footnote{We use Hermitian generators.}
\begin{equation}\label{Bianchi}
dH_3 =  \mbox{tr} R\wedge R  + \mbox{tr} F\wedge F \ .
\end{equation}
Integrated over $K3$ it yields
\begin{equation}\label{6dTadpole}
  \tfrac12 \int_{K3} dH_3 =   \chi(K3) + \tfrac12 \int_{K3} \mbox{tr} F\wedge F = 0 \ ,
\end{equation}
where 
\begin{equation}
\chi(K3)= \int_{K3}\mbox{tr} R\wedge R=24
\end{equation}
is the Euler characteristic of $K3$.
Eq.\ \eqref{6dTadpole}
implies that there has to be 
a non-trivial background gauge bundle on K3 which can either be
non-Abelian (an instanton background) or Abelian 
(a flux background). 
In the following  we consider Abelian fluxes and review
how they generate the orbifold twist in the orbifold limit
\cite{nt09}. The specific background we are interested in is the model IIIb 
 studied in ref.~\cite{ht06}.

 The Yang-Mills equation of motion plus  the Bianchi identity imply that 
the field strength $F$ is harmonic. Therefore we can expand
the Abelian background field strength as
\begin{equation}
  F =   f^I \otimes \eta_I \ , \quad I=1,\ldots,22 \ ,
\end{equation}
where $\{\eta_I\}$ is a basis of $H^2(K3,\Zbb)$ and $f^I$ are constant flux
parameters
in the Cartan subalgebra of $E_8\times E_8$.
In the following we only consider fluxes
that match with the orbifold data in the orbifold limit.

As explained in \cite{nkpt08}, the flux is related to the twist vector $V$ defined in \eqref{Vdef} and \eqref{T4twist}. $kV$ gives the non-trivial boundary conditions around the fixed locus of $\theta^k$ defined in \eqref{deftheta}. At the orbifold point, this is equivalent to making the boundary conditions trivial, but at the price of switching on a nonzero background gauge field ${\cal A}$, that has zero field strength over all the space except on the singular points. This is done by performing a non-single valued gauge transformation (see \cite{Hall:2001tn} for a simple example).  The resulting ${\cal A}$ is given by
\begin{equation}\label{matching}
   kV \equiv \oint_{\gamma_{\theta^k}} {\cal A} = \int_{D_{\theta^k}} F \:,
\end{equation}
where $D_{\theta^k}$ is a disk containing the singularity, and `$\equiv$' means 
`modulo vectors belonging to the $E_8\times E_8$ root lattice'.
In the resolved space, this integral can be written as the intersection of two two-cycles:
the two-cycle which is Poincar\'e dual to the flux $F$, that has zero size in the orbifold limit, and  a smooth cycle of K3
that in the orbifold limit contains the singularity.
The flux related to the twist is then proportional to the exceptional two-cycles (the new finite size cycle after resolution of the orbifold singularity). 
In particular, the flux related to the twist \eqref{T4twist} can be chosen as
\begin{equation}\label{fluxK3}
 F = \tfrac{1}{3} f \otimes \omega_F \ , 
\end{equation}
where
\begin{align}\label{fluxb}
 \omega_F &= \sum_{\alpha=1,3,5}\sum_{\beta=1,2,3} 
(e_{\alpha\beta}^{(1)}-e_{\alpha\beta}^{(0)}) \ ,\\
\tfrac{1}{3}f &= V\ , \quad f=(0^2,2,0^5)(2,0^7)\ .
\end{align}
$\omega_F$ is an integral two-form, written in terms of the Poincar\'e dual two-cycles.
$f$ labels an element of the Cartan subalgebra of $E_8\times E_8$,
which breaks $E_8$ to $SO(14)\times U(1)$. It is written in terms of an orthonormal basis with the convention 
tr$_{E_8\times E_8}f_1f_2=(f_1\cdot f_2)$.

The flux satisfies the tadpole cancellation condition \eqref{6dTadpole}:
\begin{equation}
 -\frac12 \int_{K3} \mbox{tr} F\wedge F = -\frac12 \, V^2 \,\int \omega_F\wedge \omega_F = -\frac12 \times 8\times (-6) = 24\ ,
\end{equation}
where we used that the intersection matrix of the 18 shrinking cycles generating the nine $\Zbb_3$ singularities $(-A_2)^{\oplus 9}$.

We now compute the massless spectrum given by this choice of flux (see \cite{Honecker:2006dt}).
The number of hypermultiplets transforming in the representation ${\bf R}$, associate to the bundle $W$, 
is given by the index (Riemann-Roch-Hirzebruch theorem \cite{Hirz})
\begin{equation}\label{chizeromd}
 \chi(W)_{K3} = \int_{K3} \mbox{ch}(W)\mbox{Td}(K3) = 2r + \mbox{ch}_2(W)\:,
\end{equation}
where ch$(W)=r+c_1(W)+$ch$_2(W)+...$ is the total Chern character of the bundle $W$, $r$ is its rank and Td$(K3)=1+\frac{1}{12}c_2(K3)+...$ 
the Todd class of the tangent bundle of K3. In our case the connection on the vector bundle is Abelian  and the representation 
is characterized by the charge $q$. In particular the number of hypermultiplets in a representation of charge $q$ is
\begin{equation}\label{mult}
 n_{\bf R} = 1 + (-3-q^2 \mbox{ch}_2(L)) = 1+3(q^2-1)\ .
\end{equation}
The flux background \eqref{fluxb}
breaks each $E_8$ to $SO(14)\times U(1)$ with the adjoint  ${\bf 248}$
representation of $E_8$ decomposing as 
\begin{equation}
 {\bf 248} = {\bf 91}_0+{\bf 1}_0+{\bf 14}_{-2}+{\bf 64}_1+{\bf {14}}_{2}+{\bf \overline{64}}_{-1}
\ ,
\end{equation}
where ${\bf 91}_0$ is the adjoint representation of  $SO(14)$.
Using \eqref{mult} one computes the multiplicities of the charged
hypermultiplets to be
\begin{equation}
 n_{(64,1)_{1,0}}=1 \ ,\qquad n_{(1,64)_{0,1}}=1 \ , \qquad n_{(14,1)_{-2,0}}=10 \ , \qquad n_{(1,14)_{0,-2}}=10 \ .
\end{equation}
Comparing with \eqref{ReducesSpHetK3}
we see that we have a match between the charged massless spectrum in the orbifold  and $K3$ background.

In  \cite{Honecker:2006dt,Louis:2011hp} it was shown that an Abelian flux also
gauges the shift symmetry of the $B$-field with respect 
to the ``fluxed'' $U(1)$. Through the St\"uckelberg mechanism, the vector boson of this $U(1)$ becomes massive
by eating one of the 20 hypermultiplets parameterizing 
the K3 moduli space. 
For the case at hand,
 the flux is switched on only along one direction in the Cartan subalgebra of $E_8\times E_8$. Hence only one combination of the two $U(1)$s is massive and the surviving gauge group is
\begin{equation}
 SO(14)\times SO(14) \times U(1)\ ,
\end{equation}
which indeed  coincides with the unbroken gauge group in the 
orbifold background  after giving a VEV to the scalars corresponding to the 
sizes of the blown up cycles.

\

One can also find fluxes which break $E_8\times E_8$ to smaller subgroups.
A set of fluxes generalizing the above example can be written as
\begin{equation}
 F= \tfrac{1}{3}\left(f_2\otimes\omega_F^{(2)} 
+ f_4\otimes\omega_F^{(4)}\right) \ ,
\end{equation}
where
\begin{equation}
\omega_F^{(2)} = \sum_{\alpha=1,3,5}\sum_{\beta=1,2,3} e_{\alpha\beta}^{(1)} 
\ , \quad
\omega_F^{(4)} = \sum_{\alpha=1,3,5}\sum_{\beta=1,2,3} e_{\alpha\beta}^{(0)}\ ,
\end{equation}
with $f_2=(0^2,2,0^5)(-1,1,0^6)$ and $f_4=(0^2,-2,0^5)(-2,0^7)$. One easily 
verifies that this flux satisfies the tadpole cancellation condition.
The unbroken group is now 
\begin{equation}\label{GgrSm6d}
 SO(14)\times U(1)\times SO(12)\times U(1)^2 \ .
\end{equation}
This group coincides with the unbroken gauge group of the second vacuum 
considered in Section~2.1, and one can check that also the massless spectra
match. 

It is instructive to write the flux in the form of the matching condition 
(\ref{matching}):
\begin{align}
\tfrac{1}{3} f_2 &= V + p_2\ , \quad p_2 = (0^8)(-1,1,0^6)\ , \\
\tfrac{1}{3} f_4 &= 2V + p_4\ , \quad p_4 = (0^2,-2,0^5)(-2,0^7)\ .
\end{align}
It is intriguing that the root vector $p_2$ is identical to the root
vector of the twisted state which acquires a VEV in the corresponding
orbifold vacuum. The root vector $p_4$ is parallel to the twist $V$.
For comparison, in the first example discussed in this section, where
both $SO(14)$ factors remain unbroken, the root vector $p_4$ is the same 
whereas $p_2 = 0$. The connection between the root vectors appearing
in the matching conditions (\ref{matching}) for fluxes and the root vectors
of twisted states acquiring VEVs has been observed in many examples, although
a deeper theoretical understanding needs further investigations.

\

The fluxes generate a D-term potential for the geometric moduli \cite{Honecker:2006dt,Louis:2011hp} which vanishes for (cf.~(\ref{Sigma}))
\begin{eqnarray}\label{Dterm6D}
   F \wedge \omega_i = 0\ ,
\end{eqnarray}
i.e., when $F$ is orthogonal to the K\"ahler form $j$ and the holomorphic 
two-form $\Omega_2$. In our case, an example of a flat direction solving \eqref{Dterm6D}, 
with $F$ given by \eqref{fluxK3}, reads
\begin{equation}\begin{aligned}
 j &= j^{(0)}+ \sum_{\alpha\beta} t^{\alpha\beta} \left(e_{\alpha\beta}^{(1)}+e_{\alpha\beta}^{(0)}\right)\ ,\\
 \Omega_2 &= \Omega_2^{(0)}+\sum_{\alpha\beta} z^{\alpha\beta} \left(e_{\alpha\beta}^{(1)}+e_{\alpha\beta}^{(0)}\right)
 	+ \left(\tfrac{2}{\nu}\sum_{\alpha\beta} t^{\alpha\beta} z^{\alpha\beta}\right) j^{(0)} \ ,
\end{aligned}\end{equation}
where $t^{\alpha\beta}$ and  $z^{\alpha\beta}$ are geometrical moduli 
fields and $j^{(0)}$ and $ \Omega_2^{(0)}$ are the K\"ahler and holomorphic two-form
at the singular point (i.e., when $t^{\alpha\beta}=z^{\alpha\beta}=0$). 
When $j$ and $\Omega_2$ have the forms defined above, the two-cycles 
$e_{\alpha\beta}^{(0)}$ and $e_{\alpha\beta}^{(1)}$ have the same size.\footnote{
The size of a two-cycle $C$ of K3 is given by vol$(C)= \sum_{a=1}^3|\int_C \omega_a|^2$, where $\{\omega_a\}$ is the orthonormal basis of $\Sigma$.}

We have switched on a flux that corresponds to the orbifold
twist. This flux generates a D-term that leaves some flat
directions. The corresponding orbifold twist also generates a D-term
leaving some flat directions. This allows us to relate the moduli on
both sides: In particular we can relate the moduli controlling the
size of $e_{\alpha\beta}^{(0)}$ and $e_{\alpha\beta}^{(1)}$ on the smooth
side with the respective scalars in the hypermultiplets
$\Phi_{0}^{\alpha\beta}$ and $\Phi_{1}^{\alpha\beta}$ in the twisted
spectrum on the orbifold side. Giving VEVs to $\Phi_{i}^{\alpha\beta}$
corresponds to making the cycles $e_{\alpha\beta}^{(i)}$ have finite
size. The D-term condition says that the size of $e_{\alpha\beta}^{(0)}$
must be equal to the size of $e_{\alpha\beta}^{(1)}$.

Hence, we see that we can go away from the orbifold point by giving a VEV to $\Phi_{i}^{\alpha\beta}$ along the flat direction. For large enough VEVs, we can match this flat direction with a flat direction in the geometric space of smooth compactifications of heterotic string theory, with a non trivial gauge background corresponding to the orbifold gauge twist.

\section{From 6D to 4D Compactifications}\label{Sec4D}

In this section we construct four-dimensional theories
by appropriately compactifying the 
six-dimensional backgrounds discussed in the previous section.
On the orbifold side, we  consider the space $T^6/\Zbb_{6-II}$
which can be seen as the two-step quotient of $T^4\times T^2$
\begin{equation}\label{orbieq}
T^6/\Zbb_{6-II} = \frac{\left(T^4/\Zbb_3\right)\times T^2}{\Zbb_2}\ .
\end{equation}
Therefore we compactify the $T^4/\Zbb_3$ orbifold discussed 
in Section~\ref{HetT4} on $T^2$  and then mod out
by a further $\Zbb_2$. This will introduce a new twisted sector
coming from the $\Zbb_2$ quotient. For a specific choice of Wilson lines 
the equivalence of this approach with the direct compactification
on the six-dimensional orbifold has been shown in 
Ref.~\cite{Buchmuller:2007qf}.

On the smooth side, we consider the singular space
\begin{equation}
Y_s= \frac{{\rm K3} \times T^2 }{\Zbb_2}\ ,
\end{equation}
and study the heterotic compactification on its resolution. 
As in $D=6$, we will show that the orbifold background can be viewed
as a singular limit of an appropriate smooth 
compactification.
Let us first discuss the orbifold theory.

\subsection{Heterotic compactification on $T^6/\Zbb_{6-II}$}

\begin{figure}[t]
\begin{center}
\includegraphics[width=10cm,height=2cm]{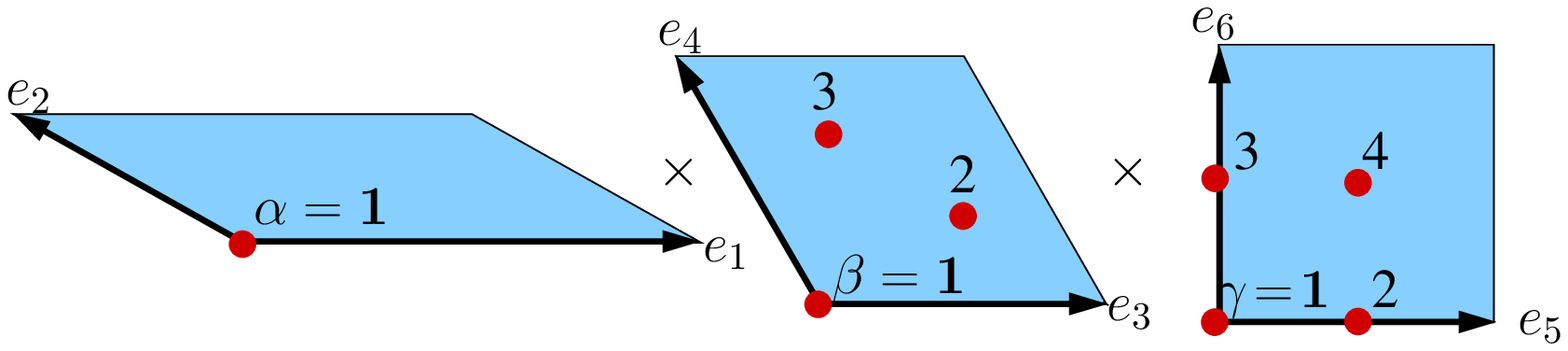}\\
\vspace*{0.5cm}
\includegraphics[width=10cm,height=2cm]{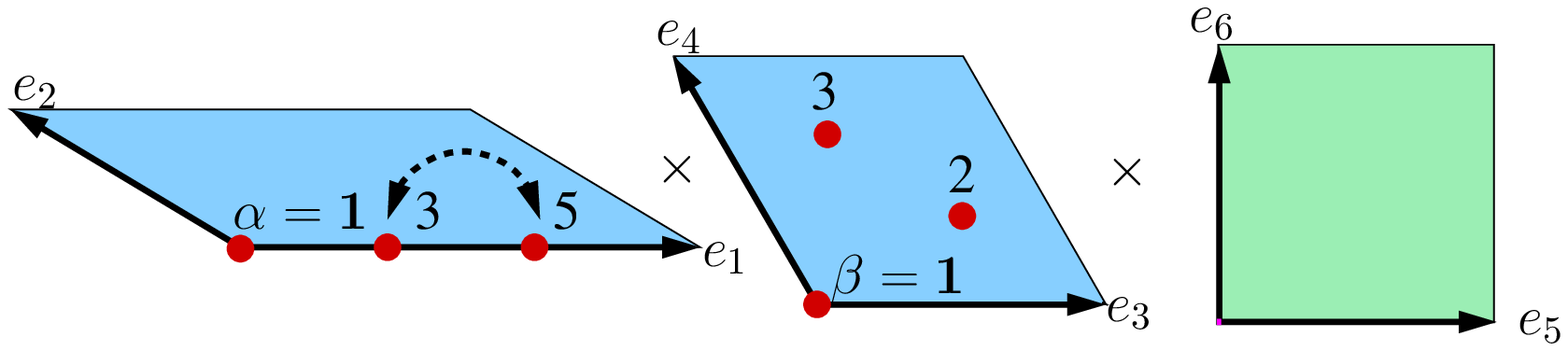}\\
\vspace*{0.5cm}
\includegraphics[width=10cm,height=2cm]{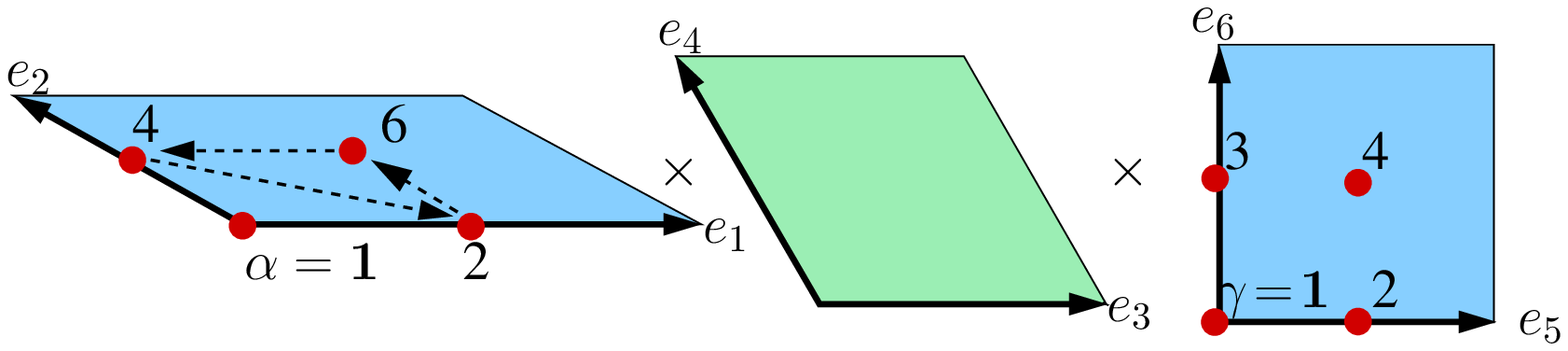}\\
\end{center}
\caption{\it Fixed points and fixed planes of\ $T^6/\Zbb_{6-II}$ orbifold for twists $\theta^k$. $\theta$: 12 fixed points (upper row); $\theta^2$: 6 fixed planes
(middle row); $\theta^3$: 8 fixed planes (lower row) (from \cite{nt09}).}
\label{fixedloci}
\end{figure}

To construct the orbifold $T^6/\Zbb_{6-II}$ we consider the six torus $T^6=\mathbb{R}^6/2\pi\Lambda$, where $\Lambda$ is the 
$\Zbb_{6-II}$ symmetric Lie algebra lattice of $G_2\times SU(3)\times SO(4)$.
The $\Zbb_{6-II}$ action $\theta$ 
on the three coordinates $z_{i=1,2,3}$ of the $T^6$ is given by
\begin{equation}\label{6orbiaction}
\theta : z_i \mapsto e^{2\pi v_6^i}\, z_i  \qquad \mbox{with} \qquad  
 v_6=\left(-\tfrac16,-\tfrac13,\tfrac12\right) \:.
\end{equation}
This orbifold has the following fixed loci \cite{Kobayashi:2004ya,
Buchmuller:2006ik} (see Fig.~\ref{fixedloci}):
\begin{itemize}
\item $\theta$ has 12 fixed points at  $(z_1,z_2,z_3)=(z_1^{\alpha=1},z_2^\beta,z_3^\gamma)$ with $\beta=1,2,3$ and $\gamma=1,...,4$, 
\item $\theta^2$ has 6 fixed planes at $(z_1,z_2)=(z_1^{\alpha},z_2^\beta)$ with $\alpha=1,3,5$ and $\beta=1,2,3$, 
where $\alpha=3$ and $\alpha=5$ are identified after the orbifold projection,
\item $\theta^3$ has 8 fixed planes at $(z_1,z_2)=(z_1^{\alpha},z_3^\gamma)$ with $\alpha=2,4,6$ and $\gamma=1,...,4$, where $\alpha=2$, $\alpha=4$ and $\alpha=6$ are identified after the orbifold projection.
\end{itemize}

An important observation is that the orbifold twist $v_6$
is the sum of a $\Zbb_3$ and a $\Zbb_2$ twist vector. The first one acts only on $(z_1,z_2)$, while the second one acts only on $(z_1,z_3)$
\begin{equation}
 v_6 = -v_3 + v_2 \ ,\qquad\mbox{where}\qquad v_3=2v_6\,, \qquad v_2 = 3v_6 \:.
\end{equation}
This allows us to reconstruct the $T^6/\Zbb_{6-II}$ orbifold in two steps 
by first acting with the $\Zbb_3$ projection and then with the $\Zbb_2$
projection as indicated in Eq.~\eqref{orbieq}.

The orbifold twist is embedded into the gauge group by the gauge twist
\begin{equation}\label{4DOrbifTwist}
 V_6 = \left( \tfrac12,\tfrac12,\tfrac13,0^5  \right) \left( \tfrac13,0^7  \right)\:.
\end{equation}
Also the gauge twist is a sum of a $\Zbb_3$ twist and a $\Zbb_2$ twist,
\begin{equation}
 V_6 = -V_3 + V_2\ , \qquad\mbox{where}\qquad V_3 = 2V_6\,, \qquad V_2 = 3V_6 \:.
\end{equation}
With this twist and vanishing Wilson lines, the unbroken gauge group is
\begin{equation}\label{4DOrbifoldGroup}
SU(2)^2 \times SO(10)\times U(1) \times [SO(14)\times U(1)]\ ,
\end{equation}
where the factor in brackets descends from the second $E_8$.
The two $U(1)$ generators are
\begin{equation}
 T_1=(0^2,2,0^5) (0^8)  \ ,\qquad T_2=(0^8) (2,0^7)\ .
\end{equation}
One can check that only one linear combination of these two $U(1)$s is non-anomalous. The anomalous and non-anomalous $U(1)$ factors are generated by
\begin{equation}
 T_{\rm non-an} = 3T_1+T_2 \, \quad\qquad T_{\rm anom}=-T_1+3T_2 \:.
\end{equation}

Massless states are characterized by the twist number $k$ and by
momentum vectors $q$ and $p$. They specify the Lorentz and gauge quantum 
numbers of the string states and are elements of the $SO(8)$ weight lattice 
and the $E_8\times E_8$ root lattice.  
States of the heterotic string are given by a direct product of the
right-moving and left-moving parts. A basis in the Hilbert space of the
quantized string is obtained by acting with  creation operators
on the ground states of untwisted and twisted sectors 
$ |q,p\rangle_{(k)} = |q + kv_6\rangle \otimes |p + kV_6\rangle $,
$k=0\ldots 4$. The vectors $\tilde{p}= p + V_6$ are called
shifted momenta. The excited states obtained by acting with creation
operators on the ground states are referred to as oscillator states.

Massless states of the untwisted and twisted sectors satisfy the 
mass equations
\begin{eqnarray}
 \tfrac{1}{8}m_\mathrm{L}^2 
 & = & 
 \tfrac{1}{2} p^2 -1+ 
 \widetilde{N}
 +\widetilde{N}^* = 0\;, \\
 \tfrac{1}{8}m_\mathrm{L}^2 
 & = & 
 \tfrac{1}{2} (p+kV_6)^2 -1+\delta c^{(k)} + 
 w_i^{(k)}\,\widetilde{N}_{i}
   +\Bar{w}_i^{(k)}\,\widetilde{N}_{i}^* = 0\;,
\end{eqnarray}
where 
\begin{equation}
 \delta c^{(k)}~=~\tfrac{1}{2}\sum_i w^{(k)}_i\,(1-w^{(k)}_i)\;,
\end{equation}
with $w^{(k)}_i\,=\,(k\,v_N)_i \mod 1$, $0<w^{(k)}_i\le1$, 
and $\Bar{w}^{(k)}_i\,=\,(-k\,v_N)_i \mod 1$, 
$0<\Bar{w}^{(k)}_i\le1$; $\widetilde{N}$, $\widetilde{N}^*$, 
 $\widetilde{N}_i$ and $\widetilde{N}_i^*$ denote various oscillator
numbers. In addition physical states have to satisfy projection conditions
which relate the vectors $q$ and $p$.

As an example consider twisted matter fields of the $T_1$ sector.
With $\delta c^{(1)} = 11/36$ they satisfy the masslessness 
condition for the shifted momenta $\tilde{p}= p + V_6$,
\begin{equation}\label{momtwist}
(p+V_6)^2~=~ \tfrac{25}{18} \ .
\end{equation}
A solution for $p+V_6$ is given by
\begin{equation}\label{example}
p+V_6 ~=~
 \left(0, 0, -\tfrac{1}{6}, 
 \text{odd}\:(\pm\tfrac{1}{2})^5\right)\left(\tfrac{1}{3},0^7\right) \ ,
\end{equation}
where ``odd $(\pm 1/2)^5$'' denotes all combinations containing an odd
number of minus signs. These states form 
a $\boldsymbol{16}$-plet of $SO(10)$. 

A standard calculation (see e.g. \cite{Schmidt:2009bd,Nilles:2011aj}) yields the massless spectrum listed in
Table~\ref{OrbifSpectrum}. The unwisted sector corresponds to $k=0$, the
twisted sectors are labeled by $k=1,2,3,4$. $\alpha,\beta,\gamma$ denote which 
fixed point locus the twisted sectors live on (see Fig.~\ref{fixedloci}). 
For example, for $\alpha=1$, $\beta=1$, $\gamma=1$, the corresponding twisted 
states are localized at the point 
$(z_1,z_2,z_3)=(z_1^{\alpha=1},z_2^{\beta=1},z_3^{\gamma=1})$. 
For fixed planes, one of the indices is blank, meaning that the twisted states
live on the whole corresponding plane. Some more explanations are needed for 
the index $\alpha$: When it takes values `$3+5$' (`$3-5$'), this means that the 
physical states are even (odd) combinations of the 6D twisted states spanning 
the planes at $\alpha=3$ and $\alpha=5$. Analogously, states labeled by 
$\alpha=2+\nu^k 4+\nu^{2k} 6$ ($k=0,1,2$, $\nu=e^{-2\pi i/3}$) are linear combinations with the phases $(1,\nu^k,\nu^{2k})$ of twisted states coming from the planes at $\alpha=2,4,6$.
\begin{table}{}
\label{OrbifSpectrum}
$$
\begin{array}{|lc|c|c|c|c|c|c|}
\hline
\mbox{Multiplet} && k & \alpha &\beta &\gamma &T_1 &T_2 \\
 \hline\hline
(1,1,1;64) && 0 & - & - & - & 0 & 1 \\
(2,2,10;1) && 0 & - & - & - & 0 & 0 \\
(1,2,16;1) && 0 & - & - & - & 1 & 0 \\
(2,1,16;1) && 0 & - & - & - & -1 & 0 \\
(1,1,1;14) && 0 & - & - & - & 0 & 2 \\
(1,1,10;1) && 0 & - & - & - & 2 & 0 \\
(2,2,1;1) && 0 & - & - & - & -2 & 0 \\
\hline \hline
(1,1,16;1) && 1 & 1 & 1,2,3 & 1,2,3,4 & -1/3 & 2/3 \\
(2,1,1;1) && 1 & 1 & 1,2,3 & 1,2,3,4 & 2/3 & 2/3 \\
(2,1,1;1) && 1 & 1 & 1,2,3 & 1,2,3,4 & 2/3 & 2/3 \\
(1,2,1;1) && 1 & 1 & 1,2,3 & 1,2,3,4 & -4/3 & 2/3 \\
\hline \hline
(1,1,1;1) && 2 & 1 & 1,2,3 & - & 4/3 & 4/3 \\
(2,2,1;1) && 2 & 1 & 1,2,3 & - & -2/3 & 4/3 \\
\hline
(1,1,1;1) && 2 & 3+5 & 1,2,3 & - & 4/3 & 4/3 \\
(2,2,1;1) && 2 & 3+5 & 1,2,3 & - & -2/3 & 4/3 \\
\hline
(1,1,1;1) && 2 & 3-5 & 1,2,3 & - & 4/3 & 4/3 \\
(1,1,10;1) && 2 & 3-5 & 1,2,3 & - & -2/3 & 4/3 \\
(1,1,1;14) && 2 & 3-5 & 1,2,3 & - & 4/3 & -2/3 \\
\hline\hline
(1,1,1;1) && 4 & 1 & 1,2,3 & - & -4/3 & -4/3 \\
(1,1,10;1) && 4 & 1 & 1,2,3 & - & 2/3 & -4/3 \\
(1,1,1;14) && 4 & 1 & 1,2,3 & - & -4/3 & 2/3 \\
\hline
(1,1,1;1) && 4 & 3+5 & 1,2,3 & - & -4/3 & -4/3 \\
(1,1,10;1) && 4 & 3+5 & 1,2,3 & - & 2/3 & -4/3 \\
(1,1,1;14) && 4 & 3+5 & 1,2,3 & - & -4/3 & 2/3 \\
\hline
(1,1,1;1) && 4 & 3-5 & 1,2,3 & - & -4/3 & -4/3 \\
(2,2,1;1) && 4 & 3-5 & 1,2,3 & - & 2/3 & -4/3 \\
\hline \hline
(1,2,1;1) && 3 & 1 & - & 1,2,3,4 & 0 & 2 \\
\hline
(1,2,1;1) && 3 & 2+4+6 & - & 1,2,3,4 & 0 & 2 \\
\hline
(1,2,1;1) && 3 & 2+\nu 4+\nu^2 6 & - & 1,2,3,4 & 0 & -2 \\
\hline
(1,2,1;14) && 3 & 2+\nu^2 4+\nu^4 6 & - & 1,2,3,4 & 0 & 0 \\
\hline
\end{array}
$$
\caption{\it Quantum numbers and localization of orbifold massless spectrum:
representations w.r.t. $SU(2)^2\times SO(10)\times SO(14)$ with $U(1)$ charges
$T_1$ and $T_2$. Untwisted states ($k=0$) are not localized. Twisted states
are localized in all planes for $k=1$; they are not localized in the
$SO(4)$ plane ($\gamma$) for $k=2,4$ and in the $SU(3)$ plane $(\beta)$ for 
$k=3$, respectively. In the $G_2$ plane $(\alpha)$ linear combinations of
states occur for the considered $\Zbb_6$ twist. }
\end{table}

\subsubsection*{Flat directions on the orbifold side}

In a 4D supersymmetric gauge theory with an anomalous $U(1)$, 
vanishing of the D-terms requires 
\begin{equation}\label{DTerm}
\begin{aligned}
 D^a &= \sum_i \phi^{\dagger}_i T^a\phi_i =0 \ , \\
D_{\rm anom} &= \sum_i q_{\rm anom}^i|\phi_i|^2 
+\frac{g M_P^2}{192\pi^2}\,\mbox{tr}(T_{\rm anom}) = 0\ .
\end{aligned}
\end{equation}
To satisfy the last equation one needs non-zero VEVs for fields that
carry opposite charges with respect to the one of the Fayet-Iliopoulos term.

The D-term conditions can be satisfied if there exists a monomial $I(\phi_i)$ 
of fields that is invariant under the non-anomalous $U(1)$'s and charged 
with respect to the anomalous $U(1)$, with sign opposite to that of 
tr$T_{\rm anom}$ (in our case tr$T_{\rm anom}=480$).
The fields entering the monomial are those that get non-zero VEV \cite{Buccella:1982nx,Casas:1987us,Cleaver:1997jb,Cleaver:1998im},
\begin{equation}
\langle \frac{\partial I}{\partial \phi_i}\rangle = c\langle \phi_i^*\rangle \ ,\end{equation}
where $\langle ...\rangle$ denotes a VEV and $c$ is some constant. Consequently,
for a monomial
\begin{equation}
I(\phi_i) = \phi_1^{n_1} \ldots \phi_l^{n_l}
\end{equation}
the various VEVs satisfy the relation
\begin{equation}
\frac{|\phi_1|}{\sqrt{n_1}} = \ldots = \frac{|\phi_l|}{\sqrt{n_l}} \ .
\end{equation}

Such flat directions indeed exist for the considered orbifold. Since we want
to realize a complete blow-up, we are interested in vacua
with non-vanishing VEVs at the locations of all fixed points and all fixed
planes, therefore involving at least 26 different fields. As we shall see, depending
on the nature of the singularity, even more than one field acquires a VEV.
Such configurations turned
out to be difficult to find and were searched for by means of a special 
computer code.  
\begin{table}[th]
$$
\begin{array}{|l|c|c|c|c|c|l|}
\hline
\mbox{Multiplet} & k & \alpha &\beta &\gamma & & \mbox{Shifted Momenta}\ 
\tilde{p}=(p+kV_6) \\
 \hline\hline
 & 1 & 1 & 1,2,3 & 1 & \tilde{p}_{1}^{\beta 1}&\left(0^2,-\frac16,-\frac12,-\frac12,-\frac12,-\frac12,-\frac12 \right)\left(\frac13,0^7\right) \\
(1,1,16;1)_{-\frac13 , \frac23} & 1 & 1 & 1,2,3 & 2 & 
\tilde{p}_{1}^{\beta 2} &\left(0^2,-\frac16,+\frac12,+\frac12,-\frac12,-\frac12,-\frac12 \right)\left(\frac13,0^7\right) \\
 & 1 & 1 & 1,2,3 & 3 & \tilde{p}_{1}^{\beta 3}&\left(0^2,-\frac16,+\frac12,-\frac12,+\frac12,+\frac12,+\frac12 \right)\left(\frac13,0^7\right) \\
 & 1 & 1 & 1,2,3 & 4 & \tilde{p}_{1}^{\beta 4}&\left(0^2,-\frac16,-\frac12,+\frac12,+\frac12,+\frac12,+\frac12 \right)\left(\frac13,0^7\right) \\
\hline \hline
(1,1,1;1)_{\frac43 , \frac43} & 2 & 1 & 1,2,3 & - &  \tilde{p}_{2;1}^\beta&\left(0,0,\frac23,0,0,0,0,0 \right)\left(\frac23,0,0^6\right) \\
\hline
(1,1,1;1)_{\frac43 , \frac43}  & 2 & 3+5 & 1,2,3 & - & \tilde{p}_{2;+}^\beta&\left(0,0,\frac23,0,0,0,0,0 \right)\left(\frac23,0,0^6\right)  \\
\hline
(1,1,1;14)_{\frac43 , -\frac23} & 2 & 3-5 & 1,2,3 & - & \tilde{p}_{2;-}^\beta&\left(0,0,\frac23,0,0,0,0,0 \right)\left(-\frac13,+1,0^6\right)  \\
\hline\hline
(1,1,1;14)_{-\frac43 , \frac23} & 4 & 1 & 1,2,3 & - &  \tilde{p}_{4;1}^\beta&\left(0,0,-\frac23,0,0,0,0,0 \right)\left(\frac13,-1,0^6\right) \\
\hline
(1,1,1;14)_{-\frac43 , \frac23}  & 4 & 3+5 & 1,2,3 & - & \tilde{p}_{4;+}^\beta&\left(0,0,-\frac23,0,0,0,0,0 \right)\left(-\frac23,0,0^6\right) \\
\hline
(1,1,1;1)_{-\frac43 , -\frac43}  & 4 & 3-5 & 1,2,3 & - & \tilde{p}_{4;-}^\beta&\left(0,0,-\frac23,0,0,0,0,0 \right)\left(-\frac23,0,0^6\right) \\
\hline \hline
(1,2,1;1)_{0,2} & 3 & 1 & - & 1,...,4 & \tilde{p}_{3;1}^\gamma&\left(+\frac12,-\frac12,0,0,0,0,0,0 \right)\left(1,0,0^6\right) \\
\hline
(1,2,1;1)_{0,2} & 3 & 2+4+6 & - & 1,...,4 & \tilde{p}_{3;+}^\gamma&\left(-\frac12,+\frac12,0,0,0,0,0,0 \right)\left(1,0,0^6\right)  \\
\hline
(1,2,1;1)_{0,-2} & 3 & 2+\nu 4+\nu^2 6 & - & 1,2 & \tilde{p}_{3;\nu}^\gamma&\left(+\frac12,-\frac12,0,0,0,0,0,0 \right)\left(-1,0,0^6\right) \\
    & 3 & 2+\nu 4+\nu^2 6 & - & 3,4 & \tilde{p}_{3;\nu}^\gamma&\left(-\frac12,+\frac12,0,0,0,0,0,0 \right)\left(-1,0,0^6\right) \\
\hline
\end{array}
$$
\caption{\it Orbifold blow-up: 42 twisted multiplets with non-zero VEVs along a
D-flat direction.}\label{FlatDirection}
\end{table}

An example involving 42 fields is listed in
Table~\ref{FlatDirection}, where the shifted momenta, i.e. the gauge quantum
numbers of the massless fields with non-zero VEVs, are given.
The corresponding monomial reads
\begin{align}\label{flatdirection}
I(\phi^{\beta\gamma}_{k;\alpha}) = 
&\left(\prod_{\beta,\gamma} \phi_{1}^{\beta\gamma}\right)^{2}  
\left(\prod_{\beta} \left(\phi_{2;1}^{\beta} 
\phi_{2;+}^{\beta}\right)^4  \left( \phi_{2;-}^\beta \phi_{4;1}^\beta \right) 
\left(\phi_{4;+}^{\beta} \phi_{4;-}^{\beta}\right)^2 \right)  \nonumber \\
&\times\left(\prod_{\gamma} \left(\phi_{3;1}^\gamma \phi_{3;+}^\gamma\right)
\left( \phi_{3;\nu}^\gamma\right)^9\right) \ ,
\end{align}
where $\phi^{\beta\gamma}_{k;\alpha}$ denotes a scalar field with quantum numbers specified
by the shifted momentum $\tilde{p}^{\beta\gamma}_{k;\alpha}$, with $k$, 
$\alpha$, $\beta$ and $\gamma$ specifying twist and location, respectively.
If the dependence on one localization index is trivial, this index is
omitted. 

Switching on these VEVs breaks the orbifold gauge group 
$SU(2)^2\times SO(10)\times U(1)\times SO(14)\times U(1)$ to the subgroup
\begin{equation}\label{OrbGaGrHig}
  SU(2)\times SU(3) \times [SO(12)]\ ,
\end{equation}
where the unbroken $SU(2)$ is one of the two $SU(2)$ factors.
In this process 
five $U(1)$ factors from the first $E_8$ and two from the second $E_8$
become massive by the Higgs mechanism.
The massless spectrum given in Table~\ref{OrbifSpectrum} can now be expressed 
in terms of 
representations of the unbroken gauge group \eqref{OrbGaGrHig}, and 
it turns out to be non-chiral  
(see Appendix \ref{HiggsedOrbifS}). Part of the spectrum will become massive 
due to superpotential couplings. Knowledge of these couplings is also needed
to find out whether the considered D-flat direction contains points in
field space where all F-terms vanish, which is required for a vacuum with
unbroken supersymmetry.

We emphasize that the complexity of the flat direction (\ref{flatdirection})
is due to the used matching conditions (\ref{matching}) and the constraints
imposed by the Bianchi identities on fluxes as discussed below. Without
these constraints, much simpler flat directions can be found, which 
leave the GUT group factor $SO(10)$ in (\ref{OrbGaGrHig}) unbroken.

\subsection{Heterotic compactification on Voisin-Borcea CY three-folds}\label{VB}
Let us now turn to the smooth side and try 
to identify a Calabi-Yau three-fold that has $T^6/\Zbb_{6-II}$ as an orbifold locus
in its moduli space.
To construct this space, we view the orbifold as a two-step quotient as in 
Eq.~\eqref{orbieq}
and use the fact that
$T^4/\Zbb_3$ can be resolved to a smooth K3, as we saw in Section~\ref{K3}. 
This leaves us with the singular space 
\begin{equation}\label{Ys}
 Y_s = \frac{K3\times T^2}{\Zbb_2}\ ,
\end{equation}
where $\Zbb_2$ acts as a $z\rightarrow -z$ involution on $T^2$ and as a non-symplectic involution on K3 \cite{Nikulin1,Nikulin2,Nikulin3}.
The action on  $T^2$ is independent of 
the resolution of the $T^4/\Zbb_3$ factor.
Thus we are left to understand 
which involution on the smooth K3 extends the
$\Zbb_2$ involution on $T^4/\Zbb_3$.
An analogous analysis for the $T^4/\Zbb_2$ limit of K3 was done in \cite{Braun:2009wh}, 
where the studied involution was the Enriques involution of K3 that extended a 
free action on $T^4/\Zbb_2$.

\subsubsection*{$\Zbb_2$ involution}

Let us first discuss the $\Zbb_2$ action on $T^4/\Zbb_3$.
It acts on the complex coordinates of $T^4$ as 
\begin{equation}\label{Z2involOrb}
(z_1,z_2) \mapsto (-z_1,z_2)\ .
\end{equation}
From Eq.~\eqref{jomegadef} we 
see that under this action the holomorphic two-form $\Omega_2$ is odd
while the K\"ahler-form $j$ is even,
 \begin{equation}
    j \mapsto +j \ ,\quad\qquad \Omega_2 \mapsto -\Omega_2 \ .
  \end{equation}
Thus the $\Zbb_2$ acts non-trivially on $T^4/\Zbb_3$, and
if we preserve this condition in the resolution we 
get a non-symplectic involution (see Appendix~\ref{AK3NSinvol} for further details). 
The fixed loci of the $\Zbb_2$ action are:
\begin{itemize}
    \item The set of points $\{(z_1,z_2)|$ $z_1=z_1^{\alpha=1}\}$. This is a $T^2/\Zbb_3$ inside  $T^4/\Zbb_3$, and it therefore has the topology of a two-sphere. 
    \item The set of points $\{(z_1,z_2)|$ $z_1=z_1^{\alpha=2}$ $\vee$ $z_1=z_1^{\alpha=4}$ $\vee$ $z_1=z_1^{\alpha=6}\}$. 
	  This set is connected and forms a $T^2$ inside  $T^4/\Zbb_3$.
\end{itemize} 

Let us now determine the action of $\Zbb_2$ on the singular points of $T^4/\Zbb_3$ and on its two-cycles.
\begin{figure}[b]
\begin{center}
\includegraphics[width=8cm]{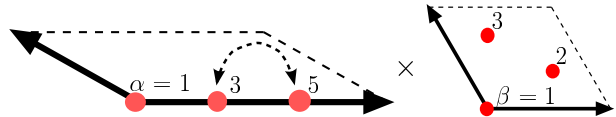}\\
\end{center}
\caption{\it $\Zbb_2$ action on the $\Zbb_3$ fixed points of $T^4$: $\alpha=1$ is fixed, while $\alpha=3,5$ are exchanged.}\label{T4Z3fig}
\end{figure}
Looking at Fig.~\ref{T4Z3fig}, we see that inverting $z_1$, one position on the $G_2$ torus of the singular points \eqref{singularpointsT4} is fixed, while the other two are exchanged:
\begin{equation}\label{FixedPtTransf}
  (z_1^{\alpha =1},z_2^{\beta}) \mapsto (z_1^{\alpha =1},z_2^{\beta}) \ ,
\qquad (z_1^{\alpha =3},z_2^{\beta}) \leftrightarrow
(z_1^{\alpha =5},z_2^{\beta}) \ ,
\end{equation}
while applying \eqref{Z2involOrb} on the four two-cycles  defined in Eq.~\eqref{four2} we obtain
\begin{equation}\label{PiABtransf}
    \Pi_1 \mapsto +\Pi_1  \ ,\qquad \Pi_2 \mapsto +\Pi_2 \ , \qquad \Theta_1\mapsto -\Theta_1 \ , \qquad \Theta_2\mapsto -\Theta_2   \ .
\end{equation}

\begin{figure}[t]
\begin{center}
\includegraphics[width=12cm]{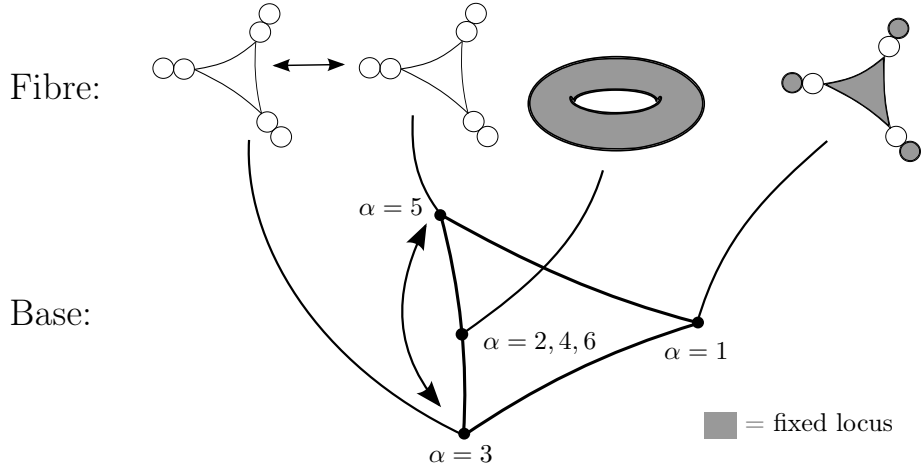}\\
\end{center}
\caption{\it $\Zbb_2$ fixed locus on the smooth K3. For a detailed description,
see Appendix~A.3.}\label{FixLocK3-2}
\end{figure}

We can resolve the orbifold singularities by
moving the K\"ahler form inside the $e^{(i)}_{\alpha\beta}$-block (defined in Section~\ref{K3}),
while keeping $\Omega_2$ fixed. In this way the cycles $e^{(i)}_{\alpha\beta}$ have non-zero size and belong to $H_{1,1}(K3)$.
The algebraic description of how to
resolve any singular point by a K\"ahler deformation is given in Appendix~\ref{AK3Blowup}. There it is also shown that
from the transformation \eqref{FixedPtTransf} of the singular points one can understand how the
shrinking cycles $e^{(i)}_{\alpha\beta}$ transform under the $\Zbb_2$ involution (see Fig.~\ref{FixLocK3-2}),  
\begin{equation}\label{trans2}
 e_{1\beta}^{(0)}\mapsto e_{1\beta}^{(0)}\ ,\qquad
 e_{1\beta}^{(1)}\mapsto e_{1\beta}^{(1)}\ , \qquad
 e_{3\beta}^{(0)}\leftrightarrow e_{5\beta}^{(0)}\ ,\qquad
 e_{3\beta}^{(1)}\leftrightarrow e_{5\beta}^{(1)}\ .
\end{equation}

The transformation of the remaining cycles is the same as in
\eqref{PiABtransf}. Therefore, we have 14 independent $\Zbb_2$ even
 two-cycles:
\begin{equation}\label{even2}
C_\ell^{K3}: \qquad
\Pi_1,\,\,\,\Pi_2,\,\,\,e^{(0)}_{1\beta},\,\,\,e^{(1)}_{1\beta},\,\,\,
e^{(0)}_{3\beta}+e^{(0)}_{5\beta},\,\,\,e^{(1)}_{3\beta}+e^{(1)}_{5\beta}
\ .
\end{equation}
Thus the rank of the even sublattice is
equal to $14$.
Furthermore,  the fixed point locus $C_{\rm fix}$ of $\Zbb_2$ acting on the smooth K3 is given by the following components:
\begin{itemize}
 \item one $T^2$ at $z_1=z_1^{\alpha=2,4,6}$, that we call $C_0^{\rm fix}$, and one $\Pbb^1$ at $z_1^{\alpha=1}$, that we call $C_4^{\rm fix}$ (these are the fixed loci of $T^4/\Zbb_3$ described above);
 \item three $\Pbb^1$s in the homology classes $e_{1\beta}^{(1)}$ that we call $C_\beta^{\rm fix}$ (see Appendix \ref{AK3Invol}).
\end{itemize}

With this information we are able to determine the non-symplectic involution
corresponding to the orbifold $\Zbb_2$. As explained in 
Appendix~\ref{AK3NSinvol}, 
a non-symplectic involution of K3 is characterized by 
its  action on the lattice of integral cycles. These involutions are
classified in terms of the even sublattice $S_+(r,a,\delta)$, where $r=$rk($S_+$) \cite{Nikulin1,Nikulin2,Nikulin3}.
The fixed locus is the disjoint union of $k$ $\mathbb{P}^1$'s and one
curve of genus $g$ where
$k = \frac12 (r-a)$ and $g = \frac12 (22-r-a)$ holds.
The fixed locus found above determines $k$ and $g$ and thus $r$ and
$a$ to be
\begin{equation}\label{kgra}
 k=4 \ , \quad g=1 \qquad \Rightarrow \qquad r=14 \ ,\quad a=6\ .
\end{equation}
Note that $r=14$ was independently derived in Eqs.~\eqref{trans2}, \eqref{even2} 
from the transformations of the K3 cycles.

So far we have kept the holomorphic two-form $\Omega_2$ fixed at the $T^4/\Zbb_3$ point. We did this to make the cycles $e_{\alpha\beta}^{(i)}$ holomorphic and 
thus algebraically describable. 
On the other hand, the most generic symmetric point in the moduli space 
corresponds to $j$ being expanded along even cycles and $\Omega_2$ along odd ones. Their generic forms are then
\begin{equation}\begin{aligned}
 j &= t_1\,\Pi_1 +  t_2\,\Pi_2 + \sum_{i=0}^1\sum_{\beta=1}^3 \left[ t_i^{1\beta} 
      e_{1\beta}^{(i)} + t_i^{3\beta} \left(e_{3\beta}^{(i)} + e_{5\beta}^{(i)}\right)\right] \:,\\
 \Omega_2 &= \zeta_1^{\Theta} \, \Theta_1 +\zeta_2^{\Theta} \, \Theta_2 + \sum_{i=0}^1\sum_{\beta=1}^3  \zeta_i^{3\beta} \left(e_{3\beta}^{(i)} - e_{5\beta}^{(i)}\right)\:.
\end{aligned}\end{equation}
Altogether we have $14$ K\"ahler moduli $t$ and $6$ complex structure moduli 
$\zeta$. Note that $\zeta_1^{\Theta}$ and $\zeta_2^{\Theta}$ are fixed in terms of the $\zeta_i^{\alpha\beta}$ by requiring $\int\Omega_2^2=0$ and $ \int\Omega_2\wedge\bar{\Omega}_2=1$.

\subsubsection*{Cycles in $Y_s$}

Once we have identified the involution that acts on K3, the
space $Y_s=(K3 \times T^2)/\Zbb_2$ in \eqref{Ys} is determined. Since
the $\Zbb_2$ involution has fixed points, $Y_s$ is singular. Before
blowing up these singularities, let us describe the finite size
cycles of $Y_s$ that are inherited
from  $K3\times T^2$. These cycles are products of K3 cycles with $T^2$ cycles that are not projected out by $\Zbb_2$. 
There are no 1-cycles (and 5-cycles) surviving the $\Zbb_2$ projection.
The surviving two-cycles are the $r=14$ even two-cycles of K3 and the
even $T^2$, i.e.\ there 
are $r+1=15$. The surviving three-cycles are the product of the $22-r$
odd two-cycles of K3 and the $2$ odd 1-cycles of $T^2$, i.e.\ there are
$2(22-r)=16$.
The four-cycles are K3 itself (that we call $R_3$ in the following to match
the notation of Ref.~\cite{nt09}) and the products of even two-cycle $C_\ell^{K3}$ of K3 (given in \eqref{even2}) with $T^2$
\begin{equation}\label{Cik3}
(C_\ell^{K3}\times T^2)/\Zbb_2\ .
\end{equation}
This is a singular four-dimensional submanifold if $C_\ell^{K3}$ has fixed points under $\Zbb_2$.

\subsubsection*{From $Y_s$ to a smooth Voisin-Borcea manifold}

\begin{figure}[t]
\begin{center}
\includegraphics[width=7cm]{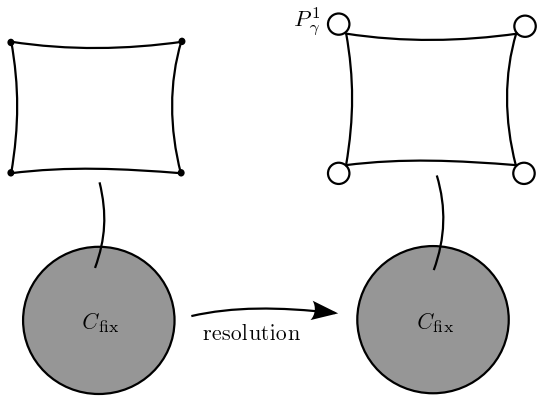}\\
\end{center}
\caption{\it Resolution of the $\mathbb{C}^2/\Zbb_2$-singularities along the $\Zbb_2$-fixed locus of K3.}\label{Z2SingResol}
\end{figure}

So far we discussed the singular space $Y_s$ and thus the next step is to
construct from it a smooth Calabi-Yau three-fold $Y$. We discuss this process keeping $r$, $a$, $k$ and $g$ generic.
We use the fact that $Y_s$ can be viewed as an elliptic fibration over $K3/\Zbb_2$ where
at a generic point of the $K3/\Zbb_2$ base the fiber is a $T^2$
\cite{mv96}.\footnote{$Y_s$ can also be understood as a K3 fibration over $T^2/\Zbb_2$.}
Over the $\Zbb_2$-fixed locus $C_{\rm fix}$ inside K3, the fiber is the
singular $T^2/\Zbb_2$ (see Fig.~\ref{Z2SingResol} on the left) with singularities being of type $A_1^{\oplus 4}$.

Now we can blow up the $A_1$ singularities to get the smooth CY three-fold $Y$. 
After the blow-up, the fiber over $C_{\rm fix}$ is a collection of five (four independent) $\mathbb{P}^1$'s intersecting like
the extended Dynkin diagram of $SO(8)$ \cite{mv96} (see Fig.~\ref{Z2SingResol} on the right).
The Hodge numbers of  $Y$ are determined by $r$ and $a$  or
alternatively by $k$ and $g$ to be \cite{Borcea,mv96}
\begin{eqnarray}
 \label{h11} h^{1,1}(Y) &=& 5+3r-2a = r+1+4(k+1) \ , \\ 
\label{h21} h^{2,1}(Y) &=& 65 -3r -2a = 1 + (20-r) +4g \ .
\end{eqnarray}
These numbers can be understood in the following way:
\begin{itemize}
  \item $h^{1,1}$ counts the number of four-cycles. We have $(r+1)$
     four-cycles determined above together with four four-cycles from the four 
blown-up $\mathbb{P}^1_{\gamma}$ ($\gamma=1,...,4$) fibered over the
$\Zbb_2$-fixed curves $C_{\rm fix}$.
Since we have $(k+1)$ such curves, we expect $4(k+1)$ new four-cycles from the 
blow-up. This matches with the expression \eqref{h11}.
  \item \label{ComplStrVBres}$h^{2,1}$ is related to the number of
     three-cycles $b_3$ via $b_3=2(h^{2,1}+1)$. Above we determined the
     number of ``bulk'' three-cycles to be  $2(22-r)$. After the blow-up,
     additional three-cycles arise as the product of the  $2g$ one-cycles in $C_g$
     and the four independent two-cycles over $C_g$, i.e., altogether
     $8g$ three-cycles.  In total we then have $b_3=2(1+h^{2,1}) =
     2(22-r)+8g$, which matches \eqref{h21}. We can alternatively understand $h^{2,1}$
   by counting complex structure moduli of $Y$. 
There is one complex structure modulus of $T^2$ and $(20-r)$ possible
complex structure deformations of $\Omega_2$.  (Recall that $\Omega_2$ can be moved only in the space of odd cycles.\footnote{The number of moduli is determined by the
    motion of a positive norm two-plane in a space with signature $(2,20-r)$, and then is equal to the
    dimension of $SO(2,20-r)/SO(2)\times SO(20-r)$.}) 
Moreover there are  $g$ complex structure deformations of the four $C_g\times \mathbb{P}^1_\gamma$ ($\gamma=1,...,4$).
\end{itemize}

Inserting \eqref{kgra} into  \eqref{h11} and \eqref{h21} we obtain
\begin{equation}
 h^{1,1}(Y) = 35 \ ,\qquad \qquad h^{2,1}(Y) = 11 \:.
\end{equation}
These are indeed the same Hodge numbers as obtained in the resolution
of $T^6/\Zbb_{6-II}$ in Ref.~\cite{nt09}. In the following we will
also see that the Voisin-Borcea manifold $Y$ is
part of the possible triangulations given in \cite{nt09}.
We shall explicitly construct the
$h^{1,1}(Y)=35$ four-cycles (divisors of $Y$) by
following the two-step blow-up procedure outlined above, and 
we shall then connect them to the ``sliding divisors'' and 
the exceptional divisors obtained by toric resolution in \cite{nt09}.

\subsubsection*{Divisors in Voisin-Borcea CY three-fold}

After the blow-up, the cycles \eqref{Cik3} give rise to smooth
four-cycles in $Y$. Let us briefly discuss  their topology that depends on which even cycle $C_\ell^{K3}$ is considered among the ones in \eqref{even2}:
\begin{itemize}
\item $C_\ell^{K3}=2\Pi_1$: $C_\ell^{K3}$ is given by two points 
on the  $T^2/\Zbb_3$ base of the elliptically fibered K3
which are exchanged by $\Zbb_2$. We call the corresponding four-cycle $R_1$. It has the topology of $T^2\times T^2$.
\item $C_\ell^{K3}=\Pi_2$: Before the blow-up, the corresponding four-cycle is given by $T^4/\Zbb_2$. We call its blow-up $R_2$. It is a K3 surface.
\item $C_\ell^{K3}=e^{(0)}_{1\beta}$: Before the blow-up the corresponding four-cycles are $(\mathbb{P}^1\times T^2)/\Zbb_2$. After the blow-up, they are called $E_{4,1\beta}$ and have the topology of an Enriques surface.
\item $C_\ell^{K3}=e^{(1)}_{1\beta}$: The corresponding four-cycles, called $E_{2,1\beta}$, have the topology of $\mathbb{P}^1\times\mathbb{P}^1$.
\item $C_\ell^{K3}=e^{(0)}_{3\beta}+e^{(0)}_{5\beta},e^{(1)}_{3\beta}+e^{(1)}_{5\beta}$: The corresponding four-cycles, called $E_{4,3\beta}$ and $E_{2,3\beta}$, have the topology of $\mathbb{P}^1\times T^2$.
\end{itemize}

To these divisors, we have to add the K3 divisor $R_3$ and the cycles that come from blowing up the $\Zbb_2$ singularities. They are given by the product 
\begin{equation}\label{Chfix}
 C_h^{\rm fix}\times \mathbb{P}^1_{\gamma}\ ,
\end{equation}
where $\gamma=1,...,4$ labels the four $\mathbb{P}^1$'s arising by the blow-up (see Fig.~\ref{Z2SingResol}), and $C_h^{\rm fix}$ are the K3 two-cycles fixed under $\Zbb_2$ shown in Fig.~\ref{FixLocK3-2}. 
We recall that for $h=0$, $C_h^{\rm fix}$ is the two-torus over $z_1=z_1^{\alpha=2,4,6}$, for $h=1,2,3$ $C_h^{\rm fix}$ are the fixed two-spheres located at $z_1=z_1^{\alpha=1}$ (in the homology class $e^{(1)}_{1\beta}$), and for $h=4$ it is the singular fiber over $z_1=z_1^{\alpha=1}$ (see also Appendix \ref{AK3} for details).
The topology of these cycles is easy to determine, as it follows from the topology of $C_h^{\rm fix}$ and \eqref{Chfix}:
\begin{itemize}
\item When $h=0$, the corresponding four-cycles, that we call $E_{3,2\gamma}$, have the topology of $T^2\times \mathbb{P}^1$.
\item When $h=\beta=1,2,3$, the four-cycles, that we call  $E_{1,\beta\gamma}$ are $\mathbb{P}^1\times\mathbb{P}^1$.
\item When $h=4$, the four-cycles called $E_{3,1\gamma}$ are $\mathbb{P}^1\times\mathbb{P}^1$.
\end{itemize}

We summarize the divisors of $Y$ in Table~\ref{TopDiv1} 
including the expected topology by 
specifying the Euler characteristic $\chi=h^{1,1}+2(1-2h^{1,0}+h^{2,0})$ and the holomorphic Euler characteristic $\chi_0=1-h^{1,0}+h^{2,0}$ 
(that are computed used the knowledge of the Hodge numbers of $T^2,
\mathbb{P}^1, $K3 and Enriques surface). 

\begin{table}{}
$$
\begin{array}{|c|c|c|c|c|}
\hline
\mbox{Divisor} & \mbox{K3 origin} & \mbox{Topology} & \chi & \chi_0 \\ 
\hline
R_{1} & 2\Pi_1 & T^2\times T^2 & 0 & 0\\ 
R_{2} & \Pi_2 & K3 & 24 & 2\\ 
R_{3} & \mbox{K3} & K3 & 24 & 2\\ 
E_{1,\beta\gamma} & C_\beta^{\rm fix} & \mathbb{P}^1\times\mathbb{P}^1 & 4 & 1\\ 
E_{2,1\beta} & e_{1\beta}^{(1)} & \mathbb{P}^1\times\mathbb{P}^1 & 4 & 1\\ 
E_{4,1\beta} & e_{1\beta}^{(0)} & \mbox{Enriques} & 12 & 1\\ 
E_{2,3\beta} & e_{3\beta}^{(1)}+e_{5\beta}^{(1)} & \mathbb{P}^1\times T^2 & 0 & 0\\ 
E_{4,3\beta} & e_{3\beta}^{(0)}+e_{5\beta}^{(0)} & \mathbb{P}^1\times T^2 & 0 & 0\\ 
E_{3,1\gamma} & C_4^{\rm fix} & \mathbb{P}^1\times\mathbb{P}^1 & 4 & 1\\ 
E_{3,2\gamma} & C_0^{\rm fix} & T^2\times\mathbb{P}^1 & 0 & 0\\ 
\hline
\end{array}
$$
\caption{\it The 35 ordinary and exceptional divisors of $Y$. In the second column we list the K3 two-cycles from which the divisors are constructed. The last three columns contain the topology of the divisors, Euler 
characteristic $\chi$ and holomorphic Euler characteristic $\chi_0$.}\label{TopDiv1}
\end{table}

Above we have denoted the constructed divisors with the same names used in \cite{nt09,lr06} for the exceptional and inherited divisors in the one-step toric resolutions of 
$T^6/\Zbb_{6-II}$. $R_1,R_2,R_3$ have been constructed in the same way and the exceptional divisors $E_{\ast,\ast\ast}$ 
have the same origin  in the toric resolution \cite{nt09,lr06}:
\begin{itemize}
\item $E_{2,1\beta},E_{4,1\beta},E_{2,3\beta},E_{4,3\beta}$ come from resolving the $\Zbb_3$ singularities (fixed planes spanning the third torus). 
In our construction they actually come from the K3 cycles obtained resolving the $\Zbb_3$ singularities; in particular, in both constructions 
$E_{2,3\beta},E_{4,3\beta}$ are combinations of cycles in $\alpha=3$ and cycles in $\alpha=5$ like in our construction.
\item $E_{3,1\gamma},E_{3,2\gamma}$ come from resolving the $\Zbb_2$ singularities (fixed planes spanning the second torus). 
In our construction they come from finite size cycles of $T^4/\Zbb_3$ fixed by $\Zbb_2$.
\item $E_{1,\beta\gamma}$ come from resolving the $\Zbb_6$ singularities (fixed points on $\alpha=1$ and spanned by $\beta,\gamma$). 
In our construction they come from K3 cycles that were shrunk before the first blow up and then were at singular points of $T^6/\Zbb_{6-II}$.
\end{itemize}

\begin{figure}[t]
\begin{center}
\includegraphics[width=4cm]{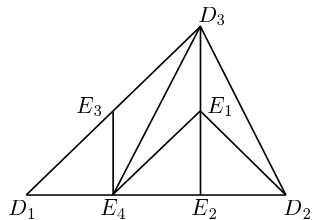}\\
\end{center}
\caption{\it Triangulation $(v)$ of $\mathbb{C}^3/\Zbb_6$ corresponding to the Voisin-Borcea resolution. The vertices correspond to divisors. Two of them intersect each other if connected by a line (from \cite{lr06,nt09}).}\label{Triangul5}
\end{figure}

$\mathbb{C}^3/\Zbb_3$ can be torically resolved by five triangulation \cite{nt09,lr06}. We now identify which of the five triangulations 
is the one corresponding to our construction. 
In our case, the cycles $E_3$ do not intersect the cycles $E_1$. In fact, they are just the product of $\mathbb{P}^1$ with $\Zbb_2$-fixed loci in K3 that do not intersect: $E_1\sim \mathbb{P}^1\times C_\beta^{\rm fix}$ and $E_3\sim \mathbb{P}^1\times C_4^{\rm fix}$ (see Table~\ref{TopDiv1}); from Fig.~\ref{FixLocK3-2} it is manifest that they do not intersect. 
There is only one triangulation that is compatible with this fact (the triangulation $(v)$ in \cite{nt09,lr06} at each fixed point, whose diagram is shown in Fig.~\ref{Triangul5}).
This allows us to use the result in \cite{nt09,lr06} for the intersection numbers of divisors and 
the characteristic classes of $Y$.\footnote{These cycles do not form an integral basis, 
in the sense that the lattice $H_4(Y,\Zbb)$ has also elements that are {\it rational} combination of them.}
The nonzero intersection numbers are:
\begin{equation}\label{IntersNumbTrIndep}
\begin{array}{lclcl}
 R_1R_2R_3 = 1 \ ,& R_2 E_{3,1\gamma}^2 = -2 \ , & R_2 E_{3,2\gamma}^2 = -6  \ ,\\
 R_3 E_{2,1\beta} E_{4,1\beta} = 1 \ , & R_3 E_{2,1\beta}^2 = -2 \ , & R_3 E_{4,1\beta}^2 = -2 \ , \\
 R_3 E_{2,3\beta} E_{4,3\beta} = 2 \ , & R_3 E_{2,3\beta}^2 = -4 \ , & R_3 E_{4,3\beta}^2 = -4 \ , \\
\end{array}
\end{equation}
where we use the abbreviation $D_iD_jD_k=\int_{Y}D_i\wedge D_j\wedge D_k$
and use the same symbol for the divisors and the Poincar\'e dual two-forms.
The intersections \eqref{IntersNumbTrIndep} are triangulation independent, i.e.\ they occur also for different toric resolution that are not related to our construction.\footnote{We can understand these intersection numbers also in our construction. $Y$ can be seen as a K3 fibration over $T^2/\Zbb_2$. Let us consider $R_3 E_{2,1\beta} E_{4,1\beta} = 1$, as an example: $R_3$ is one
point in the base $T^2/\Zbb_2$; $E_{2,1\beta}$ and $E_{4,1\beta}$ intersect on a two-cycle that is one point in the K3 fiber and wraps the base. It is then easy to see that this two-cycle intersects $R_3$ once.}
The intersection numbers corresponding to our construction are:
\begin{equation}\label{IntersNumbTrDep}
\begin{array}{lclcl}
 E_{1,\beta\gamma}E_{2,1\beta}E_{4,1\beta} = 1\ , && E_{1,\beta\gamma}E_{2,1\beta}^2= -2 \ ,
&& E_{1,\beta\gamma}^2 E_{4,1\beta}= -2\ , \\
 E_{2,1\beta}^2E_{4,1\beta} =-2 \ ,&& E_{3,1\gamma}^2E_{4,1\beta} =-2\ , && E_{1,\beta\gamma}^3 = 8\ , \\
 E_{2,1\beta}^3 = 8\ ,  && E_{3,1\gamma}^3 = 8 \ .  &&  \\
\end{array}
\end{equation}
Note that the intersection between $E_3$ and $E_1$ is zero, as it must be.

Using toric geometry, one can also compute the characteristic classes of $Y$. 
Since $Y$ is Calabi-Yau we have trivially $c_1(Y)=0$. 
For $c_2(Y)$ one has \cite{nt09,lr06}
\begin{eqnarray}\label{c2Y}
 c_2(Y) &=& -\tfrac{25}{36}\sum_{\beta\gamma}E_{1,\beta\gamma}^2 -\tfrac{5}{18}\sum_{\beta\gamma}E_{1,\beta\gamma}E_{2,1\beta} -\tfrac29 \sum_{\beta\gamma}E_{1,\beta\gamma}E_{4,1\beta} -\tfrac13 \sum_{\beta\gamma}E_{3,1\gamma}E_{4,1\beta} \nonumber \\
    && - \tfrac79 \sum_\beta (E_{2,1\beta}^2+E_{4,1\beta}^2+E_{2,3\beta}^2+E_{4,3\beta}^2) - \tfrac49 \sum_\beta(E_{2,1\beta}E_{4,1\beta}+E_{2,3\beta}E_{4,3\beta})\nonumber \\
    && -\tfrac34 \sum_\gamma (E_{3,1\gamma}^2+E_{3,2\gamma}^2)\ .
\end{eqnarray}
Once integrated on the divisor, $c_2(Y)$ gives the curvature contribution to the tadpole cancellation condition.

We can check that these data are consistent with the two-step construction: Using $c_2(Y)$ and the intersection numbers, one can derive the Euler characteristic $\chi$ and the holomorphic Euler characteristic $\chi_0$ of the divisors (see \cite{lr06} for these definitions):
\begin{equation}
 \chi(S) = \int_S c_2(S) = c_2(Y)S + S^3\ , \qquad \chi_0(S) = \tfrac{1}{12}\int_S(c_1(S)^2+c_2(S))=\tfrac{1}{12}(S^3+\chi(S))\ ,
\end{equation}
where we used $c_1(S) = -S$ and $c_2(S) = c_2(Y) + S^2$. 
The numbers that one obtains through these formulae are the same we have given in Table~\ref{TopDiv1}.

\subsection{Fluxes on the Voisin-Borcea Manifold}

In this section we determine the flux which is necessary to match the orbifold shift \eqref{4DOrbifTwist}.
We follow the same strategy as in the K3 analysis, i.e., the flux will be a linear combination of the exceptional two-forms with coefficients given
by the gauge twist of the corresponding singularity. 
Going to the intermediate step of $(K3\times T^2)/\Zbb_2$, 
the flux reduces to a flux on $K3$ plus a gauge twist on the singularity of $Y_s$. 
Following \cite{nt09}, the Ansatz for the flux is:
\begin{equation}\begin{aligned}\label{flux}
 F = &\tfrac16 \sum_{\beta\gamma} f_1^{\beta\gamma} \otimes E_{1,\beta\gamma} \\
    & +\tfrac13 \sum_\beta \left( f_2^{1\beta} \otimes E_{2,1\beta}+f_4^{1\beta} \otimes E_{4,1\beta}+f_2^{3\beta} \otimes E_{2,3\beta}+f_4^{3\beta} \otimes E_{4,3\beta} \right)\\
    & +\tfrac12 \sum_\gamma \left( f_3^{1\gamma} \otimes E_{3,1\gamma}+f_3^{2\gamma} \otimes E_{3,2\gamma} \right)\ ,
\end{aligned}\end{equation}
where the coefficients $f$  correspond to elements of the Cartan subalgebra. 
We now assume the matching conditions (\ref{matching})\cite{nt09} :
\begin{equation}\label{FluxShiftGenRel}
 \tfrac16 f_1^{\beta\gamma} \equiv V \ ,\qquad \tfrac13 f_2^{1\beta} \equiv \tfrac13 f_2^{3\beta} \equiv -\tfrac13f_4^{1\beta} \equiv -\tfrac13 f_4^{3\beta}\equiv 2V  \ ,
\qquad \tfrac12 f_3^{1\gamma} \equiv \tfrac12 f_3^{2\gamma}\equiv 3V  \ ,
\end{equation}
where $V$ is the orbifold shift given in Eq.~\eqref{4DOrbifTwist}. 
These conditions also ensure the integrality of the flux. 

If the flux coefficients $f$ were as in \eqref{FluxShiftGenRel} but with `$=$' instead of `$\equiv$', then the flux would
break $E_8\times E_8$ to the orbifold gauge group \eqref{4DOrbifoldGroup}. 
This would correspond to switching on only singlet VEVs  
in the non-Abelian factors.
On the other hand, switching on fluxes that break the gauge group to some subgroup of \eqref{4DOrbifoldGroup} corresponds, on the orbifold side, to giving VEVs to 
twisted states that are charged under the non-Abelian part of \eqref{4DOrbifoldGroup}.
This correspondence has to work 
at any singularity: A flux along the exceptional cycle $E_k$ that breaks the gauge group further should correspond to a VEV of
a twisted state in the sector $k$, that is in a representation of the non-Abelian part of the orbifold group and that leads to the same breaking.
The precise prescription is given in \cite{nt09,bn10}: The shifted momentum $\tilde{p}_{k;(i)}$ (cf.~(\ref{momtwist}))
of the twisted state $\phi_{k;(i)}$ taking a VEV, and 
localized at the singularity $(i)$, is identified with the coefficient of the flux along the exceptional divisor~$E_{k,i}$. 

In the following we  analyze the consistency conditions one has 
to impose on the flux in a smooth compactification. In searching for fluxes that
solve these conditions, we only allow for fluxes that can be related to the corresponding 
shifted momenta. In particular, they have to realize the same breaking of the gauge group.

Let us first analyze the consistency conditions imposed by the  Bianchi identities. Since the field strength $H$
of the two-form field $B$ has to be globally defined one has
\begin{equation}\label{BIY}
 \int_S dH =  -\tfrac12\int_S \mbox{tr} F\wedge F - \int_S c_2(Y) = 0\ ,
\qquad \forall\, S\ ,
\end{equation}
where $S$ is any divisor of $Y$. 
We see that \eqref{BIY} relates fluxes and geometry.
The divisors that give non-trivial conditions are $S=R_2,R_3,E_{1,\beta\gamma},E_{2,1\beta},E_{3,1\gamma},E_{4,1\beta}$:
\begin{eqnarray}
 R_2 : &&  \sum_\gamma
 \left[\left\Vert\tfrac12{f_3^{1\gamma}}\right\Vert^2+3\left\Vert\tfrac12{f_3^{2\gamma}}\right\Vert^2\right] =24 \ ,\label{R2eq}\\
 R_3 : && \sum_\beta \left[\left(\tfrac13{f_2^{1\beta}};\tfrac13{f_4^{1\beta}}\right)+ 2\left(\tfrac13{f_2^{3\beta}};\tfrac13{f_4^{3\beta}}\right) \right] =24\ , \label{R3eq}\\
 E_{1,\beta\gamma} : &&  
-4 \left\Vert\tfrac16{f_1^{\beta\gamma}}\right\Vert^2+\left\Vert\tfrac13{f_2^{1\beta}}\right\Vert^2 -\tfrac19{f_2^{1\beta}}\cdot{f_4^{1\beta}} +\tfrac19{f_1^{\beta\gamma}}\cdot{f_4^{1\beta}} = -4\ , \label{E1eq}\\
 E_{2,1\beta} : &&  -4 \left\Vert\tfrac13{f_2^{1\beta}}\right\Vert^2 
+ \tfrac29{f_2^{1\beta}}\cdot{f_4^{1\beta}}  +\sum_\gamma \tfrac16{f_1^{\beta\gamma}}\cdot\left(\tfrac23{f_2^{1\beta}}-\tfrac13{f_4^{1\beta}}\right) = -4\ , \label{E2eq}\\
 E_{3,1\gamma} : &&  -4 \left\Vert\tfrac12{f_3^{1\gamma}}\right\Vert^2 
+ \tfrac13\sum_\beta {f_3^{1\gamma}}\cdot{f_4^{1\beta}}  = -4\ , \label{E3eq}\\
 E_{4,1\beta} : &&  \left\Vert\tfrac13{f_2^{1\beta}}\right\Vert^2 
+ \sum_\gamma \left[ \left\Vert\tfrac16{f_1^{\beta\gamma}}\right\Vert^2  
-  \tfrac1{18}{f_1^{\beta\gamma}}\cdot{f_2^{1\beta}} 
+ \left\Vert\tfrac12{f_3^{1\gamma}}\right\Vert^2\right] = 12 \ ,\label{E4eq}
\end{eqnarray}
where we used \eqref{flux}, \eqref{c2Y}, the intersection numbers \eqref{IntersNumbTrIndep} and \eqref{IntersNumbTrDep},
tr$(V_1V_2) = (V_1\cdot V_2)$ and the definition 
\begin{equation}
(v;w):= \Vert v\Vert^2+\Vert w\Vert^2-v\cdot w \ .
\end{equation}
As we shall see, solving these conditions puts severe constraints 
on the fluxes. 

Eq.~\eqref{R2eq} limits the form of the coefficients of $E_{3,1\gamma}$ and $E_{3,2\gamma}$ since it 
is a sum of positive norms. If one also considers the  constraints \eqref{FluxShiftGenRel}, the minimal value
of such norms is $3/2$ which, in fact, saturates the sum. 
The possibilities for the coefficients of $f_3$ are quite restricted,
\begin{equation}
 \tfrac12{f_3} = \left( \pm \tfrac12 , \mp \tfrac12, 0^6\right) \left( \cdot\cdot\cdot \pm 1 \cdot\cdot\cdot \right)\ ,
\end{equation}
where the notation means that in the second $E_8$ the flux can only have one coefficient equal to $+1$ or $-1$, while all the others must be equal to zero.

The condition \eqref{R3eq} requires for 
$\left(\tfrac13{f_2};\tfrac13{f_4}\right)$ the minimal value 8/3, which again saturates the sum. 
Therefore we consider 
\begin{equation}\begin{aligned}
 \tfrac13{f_2} &= \left( 0,0, \tfrac23 , 0^5\right)\left( \tfrac23, 0^7\right) + \left(x_1,...,x_8\right)\left( x_9, ..., x_{16}\right)\ , \\
 \tfrac13{f_4} &= \left( 0,0, -\tfrac23 , 0^5\right)\left( -\tfrac23, 0^7\right) + \left(y_1,...,y_8\right)\left( y_9, ..., y_{16}\right)\ .
\end{aligned}\end{equation}
When all $x_i$ and $y_i$ are equal to zero, we have 
$\left(\tfrac13{f_2};\tfrac13{f_4}\right) = \tfrac83$. This value is also realized when 
$x_3$ or $x_9$ is equal to $-1$ and/or at least one of the 
$y_3, y_9$ is equal to $+1$. Because of \eqref{FluxShiftGenRel}, 
we want $(x_1,...x_8)(x_9,...,x_{16})$ and $(y_1,...,y_8)( y_9, ..., y_{16})$ 
to be elements of the $E_8\times E_8$ root lattice. Then, when $x_3=- 1$, 
one other coefficient among the $x_k$ must be equal to $\pm 1$; the same holds when $x_9$, $y_3$ and $y_9$ are different from zero.  

Remember that we are only considering fluxes which correspond to 
the orbifold data in the orbifold limit. The flux along $E_k$ can only generate
a breaking corresponding to a non-zero VEV of one twisted state in the  sector $k$. Moreover, we require that only massless twisted states acquire
non-zero VEVs.
So the flux coefficient can only be among the shifted momenta of the twisted states given in Table~\ref{OrbifSpectrum}. 
This information is crucial for the solution of the Bianchi identities. 

Consider now the Eq.~\eqref{E3eq}.
Looking at $k=3$ and $\alpha=1$ in Table~\ref{OrbifSpectrum}, 
we see that the only choices for ${f_3^{1\gamma}}$ are:
\begin{equation}\label{f31g}
 \tfrac12{f_3^{1\gamma}}= \left( \pm \tfrac12 , \mp \tfrac12, 0^6\right) \left( 1,0^7 \right)\ ,
\end{equation}
as the only twisted state is a doublet of $SU(2)$ with charge $+2$ with respect to the $U(1)$ generated by $T_2$. This is consistent with
the general solution of \eqref{R2eq}. 
On the other hand, consider $k=4$ and $\alpha=1$. All of the three states are consistent with the constraints on $f_4$ given above.
The only choice that satisfies Eq.~\eqref{E3eq}, with ${f_3^{1\gamma}}$ 
given in \eqref{f31g}, is 
\begin{equation}\label{f41b}
 \tfrac12{f_4^{1\beta}} = \left( 0^2,-\tfrac23 ,  0^5\right) \left( \tfrac13,\pm 1, 0^6 \right) \ , \qquad \forall \beta \ ,
\end{equation}
which corresponds to the state $(1,1,1;14)$ in Table~\ref{OrbifSpectrum}. 

Now we apply analogous considerations to condition \eqref{E1eq}. In the sector $k=2$ and $\alpha=1$, there are only two possible states (both consistent with the
constraints on $f_2$). In the sector  $k=1$ and $\alpha=1$ there are four 
multiplets. Using Eq.~\eqref{E1eq}, we see that the only states 
that give a solution belong to the multiplet $(1,1,16;1)$; all the other choices would correspond to fluxes that violate Eq.~\eqref{E1eq}.
Eq.~\eqref{E4eq} is automatically satisfied by this choice, while Eq.~\eqref{E2eq} is a linear combination of \eqref{E1eq} and \eqref{E4eq}. 

On the orbifold side, we  found a flat direction by giving  VEVs
to the set of twisted states of Table~\ref{FlatDirection}.
One can check that the Bianchi identities\footnote{
In particular, we see that if we take the anisotropic limit in both directions $(K3_{IIIb}\times T^2)/\Zbb_2$ and
$(K3_{IIb}\times T^2)/\Zbb_3$, the flux on the cycles that remain large (K3 cycles) is that given in models IIIb and IIb of \cite{ht06}.
Actually the flux in the first case is the one we discussed as the second choice for IIIb.} are solved by fluxes corresponding to the shifted momenta of these twisted states,
\begin{equation}\label{fluxsol}
\begin{array}{rcllcrcll}
\tfrac16{f_1^{\beta\gamma}} &=& \tilde{p}_{1}^{\beta\gamma}\ 
& \forall \beta,\gamma \ , &&
\tfrac12{f_3^{2\gamma}} &=& \tilde{p}_{3;+}^\gamma & \forall \gamma \ ,\\
  \tfrac13{f_2^{1\beta}} &=& \tilde{p}_{2;1}^\beta & \forall \beta \ , &&    
\tfrac13{f_4^{1\beta}} &=& \tilde{p}_{4;1}^\beta & \mbox{for } \forall \beta \ ,\\
  \tfrac13{f_2^{3\beta}} &=& \tilde{p}_{2;+}^\beta & \forall \beta\ ,  &&   
\tfrac13{f_4^{1\beta}} &=& \tilde{p}_{4;-}^\beta & \mbox{for } \forall \beta \ ,\\
  \tfrac12{f_3^{1\gamma}} &=& \tilde{p}_{3;1}^\gamma & \forall \gamma\ ,  &&  
\tfrac13{f_4^{3\beta}} &=& \tilde{p}_{4;+}^\beta & \forall \beta \ ;\\   
\end{array}
\end{equation}
here $\tilde{p}_{k;\alpha}^{\beta\gamma}$ are the orbifold shifted momenta of the twisted states in Table~\ref{FlatDirection}. Hence, we can indeed match 
the flux solving the Bianchi identities with the shift vectors of the
twisted states whose generic VEVs satisfy the D-term conditions \eqref{DTerm}.
 Thus the flux \eqref{fluxsol} leaves the following gauge group unbroken
\begin{equation}\label{BrokenGG}
SU(2)\times SU(3) \times U(1)^5 \times SO(12)\times U(1)^2  \ .
\end{equation}

The above flux is a linear combination of several linearly dependent generators. If we express it in terms of independent generators, it takes the form  
\begin{equation}\label{Flux7gen}
 F = \sum_{m=1}^7 f_m {\cal F}_m \ ,
\end{equation}
where the $f_m$ are the generators $T_1^{(1)},T_1^{(2)},T_1^{(3)},T_2,T_3,\tilde{T}_2,\tilde{T}_4$ of the seven $U(1)$ factors in \eqref{BrokenGG},
given in  Appendix~\ref{chirals}, and the ${\cal F}_m$ are the corresponding coefficients, which are integral two-forms living in the space spanned by the Poincar\'e dual of the exceptional divisors~$E_k$.

Using this expression we have checked that the flux also satisfies the K-theory constraints
\begin{equation}
 \sum_m c_1(L_m) =\sum_m {\cal F}_m \in H^2(T,2\Zbb)\: , 
\end{equation}
which says that the second Stiefel-Whitney class of the gauge bundle has to vanish \cite{bhw05,dg88,w85f86} (see also Ref.~\cite{W06} for a review).

\subsection{Geometric moduli vs twisted states}

Let us now determine the correspondence between the twisted states of the orbifold background and geometric moduli which govern the size of cycles 
that are blown up when moving away from the orbifold locus. 
If these states are flat directions one can blow up the orbifold completely 
and arrive at a smooth Calabi-Yau space. 

As explained in Refs.~\cite{nt09,bn10}, for a given choice of flux one can 
identify the twisted states which have non-zero VEV. The prescription is the following:
Take the flux \eqref{flux} that is a linear combination of the (Poincar\'e dual of the) exceptional divisors. Their coefficients are vectors in the weight space of $E_8\times E_8$. The twisted state in the twisted sector $k$ that takes a non-zero VEV is the one that has the shifted momentum equal to the coefficient 
$f_k$ of the exceptional divisor $E_k$.
It is then straightforward to
identify the twisted K\"ahler moduli $t_{E_k}$ using the flux \eqref{flux} with $f_k$ given by \eqref{fluxsol}.
The $t_{E_k}$ are the volumes of the 32 twisted states shown in Table \ref{match}.

\begin{table}[t]
$$
\begin{array}{|rllc|c|c|c|c|c|c|}
\hline
&&   \mbox{Multiplet} && k & \alpha &\beta &\gamma &T_1 &T_2 \\
\hline\hline t_{E_{1,\beta\gamma}}& \hookrightarrow&  (1,1,16;1) && 1 & 1 & 1,2,3 & 1,2,3,4 & -1/3 & 2/3 \\
 &&  (2,1,1;1) && 1 & 1 & 1,2,3 & 1,2,3,4 & 2/3 & 2/3 \\
 &&  (2,1,1;1) && 1 & 1 & 1,2,3 & 1,2,3,4 & 2/3 & 2/3 \\
 &&  (1,2,1;1) && 1 & 1 & 1,2,3 & 1,2,3,4 & -4/3 & 2/3 \\
\hline \hline t_{E_{2,1\beta}}&\hookrightarrow&  (1,1,1;1) && 2 & 1 & 1,2,3 & - & 4/3 & 4/3 \\
 &&  (2,2,1;1) && 2 & 1 & 1,2,3 & - & -2/3 & 4/3 \\
\hline t_{E_{2,3\beta}}&\hookrightarrow& (1,1,1;1) && 2 & 3+5 & 1,2,3 & - & 4/3 & 4/3 \\
 &&  (2,2,1;1) && 2 & 3+5 & 1,2,3 & - & -2/3 & 4/3 \\
\hline  &&  (1,1,1;1) && 2 & 3-5 & 1,2,3 & - & 4/3 & 4/3 \\
  &&  (1,1,10;1) && 2 & 3-5 & 1,2,3 & - & -2/3 & 4/3 \\
  \zeta_1^{3\beta} &\hookrightarrow&(1,1,1;14) && 2 & 3-5 & 1,2,3 & - & 4/3 & -2/3 \\
\hline \hline &&  (1,1,1;1) && 4 & 1 & 1,2,3 & - & -4/3 & -4/3 \\
 &&  (1,1,10;1) && 4 & 1 & 1,2,3 & - & 2/3 & -4/3 \\
 t_{E_{4,1\beta}}&\hookrightarrow& (1,1,1;14) && 4 & 1 & 1,2,3 & - & -4/3 & 2/3 \\
\hline  && (1,1,1;1) && 4 & 3+5 & 1,2,3 & - & -4/3 & -4/3 \\
 &&  (1,1,10;1) && 4 & 3+5 & 1,2,3 & - & 2/3 & -4/3 \\
  t_{E_{4,3\beta}}&\hookrightarrow& (1,1,1;14) && 4 & 3+5 & 1,2,3 & - & -4/3 & 2/3 \\
\hline  \zeta_2^{3\beta} &\hookrightarrow&  (1,1,1;1) && 4 & 3-5 & 1,2,3 & - & -4/3 & -4/3 \\
 &&   (2,2,1;1) && 4 & 3-5 & 1,2,3 & - & 2/3 & -4/3 \\
\hline
\hline
t_{E_{3,1\gamma}}&\hookrightarrow&  (1,2,1;1) && 3 & 1 & - & 1,2,3,4 & 0 & 2 \\
\hline   t_{E_{3,2\gamma}}&\hookrightarrow&    (1,2,1;1) && 3 & 2+4+6 & - & 1,2,3,4 & 0 & 2 \\
\hline \zeta^{2\gamma} &\hookrightarrow&  (1,2,1;1) && 3 & 2+\nu 4+\nu^26 & - & 1,2,3,4 & 0 & -2 \\
\hline  && (1,2,1;14) && 3 & 2+\nu^24+\nu^46 & - & 1,2,3,4 & 0 & 0 \\
\hline

\end{array}
$$
\caption{\it Geometric moduli among the twisted states
of the massless orbifold spectrum. 32 states are identified as volume moduli
$t_{E_k}$ of the corresponding exceptional divisors $E_k$; $\zeta^{3\beta}_1$,
$\zeta^{3\beta}_2$ and $\zeta^{2\gamma}_1$ denote 10 complex structure
moduli.}\label{match}
\end{table}

In addition we also want to identify the complex structure moduli. 
In a smooth background
with Abelian fluxes the complex structure moduli are not fixed. 
Therefore we look for flat directions 
among the twisted states on the orbifold side. 
In our example we have 10 ``twisted'' complex structure moduli. 
From our treatment of the resolution of the 
$T^6/\mathbb{Z}_{6-II}$ orbifold through two steps,
we know that six of them are the complex structure moduli of K3 that survive the $\mathbb{Z}_2$ projection, and the remaining four come from the complex structure deformations
of the four-cycles arising after the blow-up of the $\mathbb{Z}_2$ singularities (see discussion below Eq.~\eqref{h21}).
Since the holomorphic two-form K3 is odd under $\mathbb{Z}_2$
its deformation must be chosen from odd combinations of the twisted sectors corresponding to $\alpha=3$ and $\alpha=5$. These deformations are related to the twisted geometric moduli of the resolved $T^4/\mathbb{Z}_3$ and  hence, 
they must be 
in the $k=2$ and $k=4$ sectors in Table~\ref{OrbifSpectrum},
labeled by $\alpha=3-5$.

The remaining four twisted complex structure moduli must be in the $k=3$ sector. We can identify them by considering the orbifold
$T^6/\mathbb{Z}_{6-II}$ as $\left(T^4/\mathbb{Z}_2\times T^2\right)/\mathbb{Z}_3$. In this case $T^4/\mathbb{Z}_2$ is blown up to
a smooth $\widetilde{K3}$ and one should find a $\mathbb{Z}_3$
symmetry of 
$\widetilde{K3}$ to take 
the last quotient. This is more difficult than finding an involution and 
we will not study this procedure in detail. On the other hand, by analogy 
with the $T^4/\Zbb_3$ case, we can make an educated guess.
In fact the holomorphic two-form $\Omega_{\widetilde{K3}}$ should
transform with a phase under the $\mathbb{Z}_3$ and therefore 
its deformation should exhibit the same phase. 
The states with this property are
labeled by $\alpha=2+\nu4+\nu^26$ and $\alpha= 2+\nu^24+\nu^46$ in Table~\ref{OrbifSpectrum}. 
Depending on the choice of the complex structure it will be one of the 
two states.\footnote{With our choice of the orbifold action given in \eqref{6orbiaction}
the $T^4/\mathbb{Z}_2$ in question is along $z_1,z_3$. In the singular limit, the holomorphic two-form is $\Omega_{\tilde{K3}}\sim dz_1\wedge dz_3$ and transforms
under $\mathbb{Z}_3$ as $\Omega_{\tilde{K3}}\mapsto e^{2\pi i/3}\Omega_{\tilde{K3}}$. This is the transformation of the states labeled by $\alpha=2+\nu 4+\nu^2 6$, where 
$\nu=e^{-2\pi i/3}$.
}
The matching of the geometric moduli with the twisted states is summarized in Table~\ref{match}.

\subsection{The spectrum on the smooth Calabi-Yau}\label{smoothS}

The flux given in Eq.~\eqref{fluxsol} induces the following breaking of
the gauge group 
\begin{equation}\label{4Dbrokengroup}
E_8\times E_8\quad\to\quad
SU(2)\times SU(3) \times U(1)^5 \times SO(12)\times U(1)^2 \ .
\end{equation}
The adjoint representations of the two $E_8$ factors decompose according to
\begin{equation}\begin{aligned}
 {\bf 248}^{(1)} &= ({\bf 1},{\bf 3}) \oplus ({\bf 8},{\bf 1})
\oplus ({\bf 1},{\bf 1})^5 
 \oplus ({\bf R^{(1)}}\oplus {\bf \bar{R}^{(1)}})\ ,\\
 {\bf 248}^{(2)} &= {\bf 66}  \oplus {\bf 1} \oplus {\bf 1} \oplus ({\bf R^{(2)}}\oplus {\bf \bar{R}^{(2)}}) \ ,
\end{aligned}\end{equation}
where the entries in the first line correspond to $SU(2)\times SU(3)$
while the second decomposition is with respect to $SO(12)$.
${\bf R^{(1)}}$ and ${\bf R^{(2)}}$ denote reducible
representation of the unbroken gauge group. The irreducible representations in ${\bf R^{(1)}}$
and in ${\bf R^{(2)}}$ are listed in Appendix~\ref{chirals} (Tables \ref{firstE8} and \ref{secondE8}).
Since they are charged under the fluxed $U(1)$s, there can be chiral states in these representation.
The chiral multiplicity is given by an index theorem (see for example \cite{GSW}) . Following \cite{Nibbelink:2007pn,nt09}, we
can use the ``number operator''
\begin{equation}\label{Nop}
 N = \tfrac{1}{6}\int_X\left( F\wedge F\wedge F + \tfrac12 F\wedge c_2(X) \right)\ ,
\end{equation}
to compute the spectrum. 
The eigenspaces of $N$ are the representations of the unbroken gauge group and 
the corresponding eigenvalues give the multiplicity of these representations.
Note that tracing the expression \eqref{Nop} over the representation ${\bf R}$ gives the index of the Dirac operator index$_{\bf R}\not\!\! D$ in that representation.

Inserting the expression for the flux \eqref{flux} into \eqref{Nop}, one obtains:
\begin{equation}\begin{aligned}
 N =&   \tfrac43\sum_{\beta\gamma}(V_1^{\beta\gamma})^3 +\tfrac43\sum_{\beta}(V_2^{1\beta})^3 +\tfrac43\sum_{\gamma}(V_3^{1\beta\gamma})^3  - \sum_{\beta\gamma}V_1^{\beta\gamma}(V_2^{1\beta})^2
	  - \sum_{\beta\gamma}(V_1^{\beta\gamma})^2V_4^{1\beta} \\
  & - \sum_{\beta}(V_2^{1\beta})^2V_4^{1\beta} + \sum_{\beta\gamma}V_1^{\beta\gamma} V_2^{1\beta}V_4^{1\beta} - \sum_{\beta\gamma}(V_3^{1\gamma})^2V_4^{1\beta} \\
  & -\tfrac13 \sum_{\beta\gamma} V_1^{\beta\gamma} -\tfrac13 \sum_{\beta} V_2^{1\beta} -\tfrac13 \sum_{\gamma} V_3^{1\gamma} + \sum_{\beta}V_4^{1\beta}\ . 
\end{aligned}\end{equation}
The $V_i^{\beta\gamma}$ are elements of the Cartan subalgebra  of $E_8$.
They act on an element of the adjoint representation of $E_8$ labeled by the root $\alpha$ as
\begin{equation}
 [V_i^{\beta\gamma},v_\alpha]= (\alpha\cdot V_i^{\beta\gamma})\, v_\alpha\ .
\end{equation}
Then, if we know the roots corresponding to the decompositions of the adjoint of $E_8\times E_8$, 
we are able to compute the multiplicity of the given representation. 
We find
\begin{equation}
 V_1^{\beta\gamma} = \tfrac16{f_1^{\beta\gamma}}\ , \qquad 
V_{2}^{1\beta} = \tfrac13{f_{2}^{1\beta}}\ , \qquad 
V_3^{1\gamma} = \tfrac12{f_3^{1\gamma}}\ ,
\qquad V_4^{1\beta} = \tfrac13{f_4^{1\beta}}\ .
\end{equation}
The $V_i^{\beta\gamma}$s can be written in terms of the generators of the $U(1)$ factors in the unbroken gauge group. Then the eigenvalues
of the number operator are given in terms of the charges 
of the representation ${\bf R}_i$ with respect to the $U(1)$ factors.
The detailed computation of the chiral spectrum is given in Appendix~\ref{chirals}. 

Switching on the flux \eqref{Flux7gen} renders all the seven $U(1)$ factors in \eqref{4Dbrokengroup} massive with a 
St\"uckelberg induced  mass matrix (see e.g. \cite{W06} for a review),
\begin{equation}
 M_{mn}^2 = \sum_j\tfrac{1}{\alpha_j} {\cal M}_{jm}{\cal M}_{jn}\ ,
\end{equation}
where $\alpha_j$ and ${\cal M}_{jm}$ appear in the axion part of the 4D effective action,
\begin{equation}
 S\ \subset\ \sum_{j=0}^{h^{1,1}}\,
\int_{\mathbb{R}^{1,3}}b_j^{2}\wedge \sum_m {\cal M}_{jm} F_m 
+ \alpha_j \int_{\mathbb{R}^{1,3}}db_j^{2}\wedge *db_j^{2} \ .
\end{equation}
$F_m$ is the field strength of an unbroken $U(1)$ factor and 
in our case $m$ takes the values $m=1,\ldots,7$. 
$b_j^{2}$ are the 4D axionic two-forms coming from the expansion of $B_6$ along the harmonic four-form of the Calabi-Yau, plus the universal two-form $B_2$ along
the 4D space-time. (Dualizing these two forms one gets the axionic scalars). 
We neglect the couplings ${\cal M}_{0m}$ related to the universal axion
and only consider
\begin{equation}
 {\cal M}_{jm} = \sum_{n=1}^7 (f_m\cdot f_n) \int_Y {\cal F}_n \wedge \eta_j\ ,
\end{equation}
where $\{\eta_j\}$ is a basis of $H^4(Y)$, $f_n$ are 
elements of the Cartan subalgebra of $E_8\times E_8$ and 
${\cal F}_n\in H^2(Y,\Zbb)$ is such that the flux is given by \eqref{Flux7gen}.
One observes that with our choice of flux the matrix ${\cal M}_{jm}$ has maximal rank 
rk$({\cal M}_{jm})=7$. 
This implies that all the seven $U(1)$ factors become massive due to the presence of the flux and the unbroken gauge group reduces to
\begin{equation}\label{4DUnbrGaugeGrSmooth}
 SU(2)\times SU(3)\times SO(12)\ .
\end{equation}
This is indeed the same gauge group obtained on the orbifold side on the Higgs branch.

Note that the spectrum given in Tables \ref{firstE8} and \ref{secondE8} is not chiral with respect to the surviving gauge group \eqref{4DUnbrGaugeGrSmooth}.
This makes the identification of the states more difficult than in the 6D case.
To match the two spectra, we still have to consider the charges under
the massive $U(1)$s. This forces us to redefine the states on the
orbifold side, using the singlets that acquire a  VEV (see \cite{nkpt08}).

\subsubsection*{Matching with the Higgsed orbifold spectrum}

The spectrum on the smooth side does not match with the one at the orbifold point. In fact a smooth compactification corresponds to switching on VEVs for the twisted states which correspond to the geometric moduli.
This breaks the gauge group at the orbifold locus to the subgroup which is visible 
on the smooth side via the Higgs mechanism.  Some of the chiral multiplets 
are eaten by the massive gauge bosons. Moreover, to achieve 
a match of the $U(1)$ charges, it is necessary 
to redefine the fields on the orbifold side as we did in Eq.~\eqref{redefT}. 
In the $T^6/\mathbb{Z}_6$ case this is slightly more difficult than for the 
$T^6/\Zbb_3$ orbifolds discussed in~\cite{nkpt08,Blaszczyk:2011ig}. 
The reason is 
that states in different twisted sectors can mix 
due to the fact that for $T^6/\mathbb{Z}_6$
the singular loci intersect each other.
\begin{table}[htb]
$$
\begin{array}{|c|c|ccccccc|}
\hline
&  & q_1^{(1)} & q_1^{(2)} & q_1^{(3)} & q_2 & q_3 & \tilde{q}_2 & \tilde{q}_4 \\\hline
1 &(3,1;1) & -13/3 & 5/3 & 5/3 & 0 & -1/3 & 2/3 & 1/3 \\
2 &(3,1;1) & 5/3 & -1/3 & 5/3 & 0 & -1/3 & 2/3 & 1/3 \\
3 &(\bar{3},1;1) & 5/3 & 5/3 & 5/3 & 0 & -1/3 & 2/3 & 1/3\\
4 &(\bar{3},1;1) & 5/3 & 5/3 & -1/3 & 0 & -1/3 & 2/3 & 1/3\\
5 &(1,2;1) & -1/3 & -1/3 & -1/3 & 0 & 2/3 & 2/3 & 1/3\\
6 &(1,2;1) & -1/3 & -1/3 & -1/3 & 0 & 2/3 & 2/3 & 1/3\\
7 &(1,1;1) & 2/3 & 5/3 & -1/3 & 0 & -1/3 & 2/3 & 1/3\\
8 &(1,1;1) & 5/3 & 2/3 & -1/3 & 0 & -1/3 & 2/3 & 1/3\\
9 &(1,1;1) & 2/3 & 2/3 & 2/3 & 1 & -4/3 & 2/3 & 1/3\\
10 &(1,1;1) & 2/3 & 2/3 & 2/3 & -1 & -4/3 & 2/3 & 1/3\\
11 &(1,1;1) & -1/3 & -1/3 & 2/3 & 0 & -1/3 & 2/3 & 1/3\\
12 &(1,1;1) & -13/3 & -13/3 & 5/3 & 0 & -1/3 & 2/3 & 1/3\\\hline
\end{array}
$$
\caption{\it The 22 states of the twisted sector $k=1$; 
states from the representation $(1,1,16;1)$ acquire a VEV, e.g. state \#12.}\label{k1a1b1twsts}
\end{table}
We have been able to find field redefinition which match 
the charged states on both sides.
For the singlets however, the possible redefinitions are less constrained 
and therefore we do not attempt to match them in this paper.
We do not present the complete computation here
but as a representative example go through the sector $k=1$ with $\alpha=1$, $\beta=1$, $\gamma=1$. 
The spectrum after higgsing is given in Table~\ref{k1a1b1twsts}.
\begin{table}[t]
$$
\begin{array}{|c|c|ccccccc|}
\hline
 && q_1^{(1)} & q_1^{(2)} & q_1^{(3)} & q_2 & q_3 & \tilde{q}_2 & \tilde{q}_4 \\
\hline
1& (3,1;1) & 4 & 4 & -2 & 0 & -2 & 0 & 0 \\
2& (3,1;1) & -6 & -6 & 6 & 0 & 0 & 0 & 0 \\
3& (\bar{3},1;1) & -6 & 0 & 6 & 0 & 0 & 0 & 0\\
4& (\bar{3},1;1) & -6 & 0 & 0 & 0 & 0 & 0 & 0 \\
5& (1,2;1) & -8 & -2 & 4 & 0 & 0 & 0 & 0 \\
6& (1,2;1) & -8 & -2 & 4 & 0 & 0 & 0 & 0 \\
7& (1,1;1) & 0 & 0 & 0 & 0 & 0 & 0 & 0 \\
8& (1,1;1) & -6 & 6 & 0 & 0 & 0 & 0 & 0 \\
9& (1,1;1) & -7 & -1 & 5 & 1 & -1 & 0 & 0 \\
10& (1,1;1) & -7 & -1 & 5 & 1 & 1 & 0 & 0 \\
11& (1,1;1) & 4 & -2 & 4 & 0 & -2 & 0 & 0 \\
12& (1,1;1) & 4 & -2 & -2 & 0 & -2 & 0 & 0 \\
\hline\end{array}
$$
\caption{\it The 22 states on the smooth side, which correspond to the states
of the twisted sector $k=1$.}\label{k1a1b1sm}
\end{table}
The charges refer to the $U(1)$s along which the flux is switched on in the smooth case. The field that gets non-zero VEV 
is \#12, the last one in Table~\ref{k1a1b1twsts}, which corresponds to the one in the first line of Table~\ref{FlatDirection}. We call it $\Phi_1$. It 
has the following shifted vector:
\begin{equation}\label{w11latvec}
 \tilde{p}_{1,1} = \left(0^2,-\tfrac16,-\tfrac12,-\tfrac12,-\tfrac12,-\tfrac12,-\tfrac12 \right)\left(\tfrac13,0^7\right)\:.
\end{equation}

To redefine the fields in the sector $k=1$, $\alpha=1$, $\beta=1$, $\gamma=1$, the field $\Phi_1$ is not enough to get the right match with the spectrum on the smooth side. We need to use fields from other twisted sectors that live on subspaces intersecting the point-like singularity where the states \eqref{k1a1b1twsts} live. These are the twisted states that take non-zero VEV in the other sector after the blow-up.
For the chosen twisted sector, in addition to the field $\Phi_1$ we need to consider also the field $\Phi_2$ from the sector $k=2$, $\alpha=1$, $\beta=1$, with shifted vector
\begin{equation}
\tilde{p}_{21}=\left(0,0,\tfrac23,0,0,0,0,0 \right)\left(\tfrac23,0,0^6\right)\ .
\end{equation}
In particular, we redefine the first and the last two states in Table~\ref{k1a1b1twsts} multiplying them by $\Phi_1\Phi_2^{-1}$ and the other states by $\Phi_1^{-1}$. The charges of the new fields are given in Table~\ref{k1a1b1sm}.
We end up with fields that now are charged only under one $E_8$ and can 
therefore be matched with the fields in the smooth sector.
The states \#2,3,4,8 have a complex conjugate partner (with respect to the massive $U(1)$s) coming from other $k=1$ sectors, and then decouple from the spectrum.
The state \#7 is the geometric modulus corresponding to the size of the blown up four-cycle.
The other states can be recognized to be among the states in Table~\ref{k1a1b1sm}.
To have the right multiplicities, one has to consider the complete orbifold higgsed spectrum. For example, states like state \#1 in Table~\ref{k1a1b1sm} also appear in the sectors $k=1$, $\alpha=1$, $\beta=1,2,3$, $\gamma=1,2$, with total multiplicity $6$. It also appears with opposite chirality (with respect to the massive $U(1)$s), in the untwisted spectrum with multiplicity $1$, and in the twisted $k=4$, $\alpha=1$, $\beta=1,2,3$ with multiplicity $3$. The total number of chiral states in this representation is $n=2$, that matches with the multiplicity given in the Table~\ref{firstE8}. The same happens for all the fields in the non-trivial representations of the unbroken group.

\section{Conclusions}

We have studied the relation between a specific orbifold background 
and a smooth Calabi-Yau compactification which contains the orbifold background 
as a singular locus in its moduli space. 
Specifically, we considered the orbifold 
$T^6/\Zbb_{6-II} = (T^4/\Zbb_3\times T^2)/\Zbb_2$ for which we performed 
a two-step resolution. 
As a first intermediate step we resolved $T^4/\Zbb_3$ to a smooth $K3$ 
surface. For the compactification on K3 to six dimensions 
we discussed the line bundle corresponding to the chosen $\Zbb_3$ orbifold 
gauge twist. We also
identified the twisted states which correspond to the geometric $K3$ moduli.
After giving VEVs to these states, the orbifold gauge group and its 
massless spectrum  
matched the gauge group and spectrum of the smooth compactification.

In the second step we compactified to four dimensions and studied the 
resolution of the singular space $Y_s=(K3\times T^2)/\Zbb_2$.
In particular, finding how the $\Zbb_2$ involution acts on the K3 factor, we
were led to consider 
heterotic compactifications on a specific smooth Voisin-Borcea manifold. 
This manifold is a K3 fibration and in the 
limit of a large $\mathbb{P}^1$-base one recovers the 6D compactifications on 
K3. The considered Voisin-Borcea manifold has a specific  intersection form which can
be compared with the toric resolutions studied in 
\cite{nt09}. The latter come from different triangulations 
that one can perform at each singularity of $T^6/\Zbb_{6-II}$. 
With the help of this comparison 
we were able to identify 
which triangulation corresponds to our specific 
Voisin-Borcea manifold.

In heterotic compactifications it is necessary to also identify 
the gauge bundle which solves the Bianchi identities. 
Here we considered non-trivial line bundles (fluxes) on the Voisin-Borcea 
manifold that in the orbifold limit reproduce the gauge twists up
to vectors of the $E_8\times E_8$ root lattice.
It turned out that the fluxes are severely constrained by the Bianchi 
identities.  
In particular, we found that some flux on the K3 fiber allowed in the 6D
background, no longer could be turned on. The resulting low energy gauge group
is small, even smaller than the standard model gauge group. 
We could show that for the unbroken group the charged massless states match.
However, the 
massless spectrum is non-chiral with respect to the unbroken gauge group. 
Hence, we do not obtain the phenomenologically interesting orbifold background
as limit of a related smooth CY compactification.

This result, in particular the breaking of the phenomenologically
interesting $SO(10)$ GUT group down to $SU(3)$, came as a surprise, 
and it is clearly important to understand its origin. 
The matching conditions between fluxes 
and orbifold gauge twists involve $E_8\times E_8$ root vectors which are 
related to the shifted momenta of twisted orbifold states acquiring non-zero 
VEVs. We only found solutions of the Bianchi identities where
${\bf 16}$-plets of $SO(10)$ obtain VEVs. This immediately implies a
strong breaking of the orbifold gauge group. It is conceivable that we
missed interesting solutions or that new 
solutions appear once discrete Wilson lines are turned on. A further
possibility is that the employed matching conditions, which so far represent a
conjecture, are too restrictive. 
In \cite{nt09}, the MSSM gauge group was found, but only at the expense of
assigning VEVs to massive twisted states which cannot represent moduli. 
For the toy model with standard embedding in \cite{bn10}, 
a complete blow up was achieved by giving VEVs to massless twisted states
only. However, the unbroken group does not contain the standard model gauge
group and the massless spectrum is non-chiral.
Clearly, further studies are needed to clarify the situation, and one
may also consider the case of partial blow-up.

Our study also illustrates some conceptual problems which one faces in attempts
to embed the standard model into string theory. In orbifold compactifications
one constructs massless states of the free string at a particular point in 
moduli space. Information about the couplings of twisted states located at
different singularities is only contained in the D-terms of the unbroken 4D 
gauge group. Some information on Yukawa couplings is provided by string
selection rules, but the superpotential is essentially unknown, and therefore
one cannot rigorously check the vanishing of the F-terms. Since realistic
models require to go away from the orbifold point by assigning VEVs to
twisted states, one therefore cannot be sure whether the resulting gauge group 
and spectrum really correspond to a string vacuum.

On the other hand, in Calabi-Yau compactifications one starts from the
supergravity approximation in ten dimensions. The geometry of the compact
space then provides valuable information on possible breakings of the 
$E_8\times E_8$ gauge group and the related massless spectrum via
the Bianchi identities. However,
since one is dealing with supergravity rather than string theory, the
singular orbifold limit cannot be performed and one can only compare
gauge group and massless spectrum with an orbifold construction where part of
the gauge group is higgsed. There are intriguing relations between
CY fluxes and orbifold twisted states, but the precise matching conditions
are presently unclear.  

Embedding the (supersymmetric) standard model into string theory is a
challenging problem, even before addressing the question of supersymmetry
breaking. In the long run, the matching between orbifold constructions and
supergravity compactifications on Calabi-Yau manifolds may become a useful
tool. The orbifold constructions can provide some intuition to identify 
interesting 
geometries and fluxes, which yield the wanted chiral massless spectrum. 
The corresponding smooth compactifications may then help to understand the 
dynamical stabilization of the ground state.

\newpage

\subsection*{Acknowledgments}

We would like to thank Martin Schasny for helpful discussions, and Stefan Groot Nibbelink  and Patrick
Vaudrevange for correspondence and a careful reading of the manuscript. This work has been 
supported by the German Science Foundation (DFG) within the Collaborative 
Research Center 676 ``Particles, Strings and the Early Universe'' and 
the Research Training Group 1670.

\vskip 2cm

\appendix

\noindent
{\bf\Large Appendices}

\section{Geometric details}\label{AK3}

In this appendix we supplement the main text 
with additional material about K3. Let us start by discussing 
non-symplectic involutions on K3 surfaces as used in Section~\ref{VB}.

\subsection{Non-symplectic involutions on K3 surfaces}\label{AK3NSinvol}

A holomorphic involution of K3 is called non-symplectic when it reverses the sign of the holomorphic
two-form $\Omega$. These involutions have been studied and classified by
Nikulin \cite{Nikulin1,Nikulin2,Nikulin3}, and we review here some of their properties.

The non-symplectic involutions $\sigma$ of K3 are classified in terms
of three parameters: $(r,a,\delta)$, which characterize
the lattice $S_+$ of even cycles which is  a sublattice of the Picard
lattice (as $\Omega_2\mapsto -\Omega_2$).\footnote{The Picard lattice is the orthogonal complement of $\Omega_2$ in $H_2(K3,\Zbb)$.}
$r$ is the
rank of $S_+$ while $a$ is defined by the expression
$S_+^*/S_+=(\Zbb_2)^a$.\footnote{
For each lattice $S$, there exists a dual lattice $S^*=$Hom$(S,\Zbb)$. Each element $\beta\in S^*$ can be represented
by an element $b \in S\otimes \mathbb{R}$ by identifying $\beta(\cdot)=(b,\cdot)$, where $(\cdot,\cdot)$ is the intersection
bilinear form on $S$. It can be shown that for $S$ to be the even lattice under an involution on K3, it must satisfy $S^*/S=(\Zbb_2)^a$
for some $a$ (i.e. it must be a 2-elementary lattice).
} 
$\delta$ can take two values: $\delta=0$ if $\forall (x,\cdot)\in
S_+^*$ $(x,x)\in \Zbb$, $\delta=1$ otherwise.

The triplet of parameters $(r,a,\delta)$ encodes information about the fixed point set in K3. If $(r,a,\delta)\not=(10,10,0)$ and $(r,a,\delta)\not=(10,8,0)$, then
the fixed point set is given by
\begin{equation}
 \bigcup_{i=1}^k\, \mathbb{P}^1_i  \ \cup \ C_g\ ,
\end{equation}
i.e., it is a  disjoint union of $k$ rational curves (the
$\mathbb{P}^1$s) and one curve $C_g$ with genus $g$. $k$~and $g$ are
in turn  determined by $r$ and $a$ via
\begin{equation}\label{kg}
  k = \tfrac12 (r-a)\ , \quad \qquad g = \tfrac12 (22-r-a) \:.
\end{equation}
When $(r,a,\delta)=(10,10,0)$ the involution acts freely (called Enriques involution), while if $(r,a,\delta)=(10,8,0)$ the fixed point set
is a disjoint union of two elliptic curves (two two-tori).

\subsection{Blowing up the $A_2$-singularities}\label{AK3Blowup}

Let us now consider the geometric transition from $T^4/\Zbb_3$ to a smooth K3 from an algebraic point of view.
The nine $A_2$-singularities can be resolved by rotating the K\"ahler
form $j$ into the space of  $A_2$-cycles such that they are no
longer orthogonal to the $\Sigma$ plane and thus have finite size.

From an algebraic point of view, the blow-up is described as follows
(see \cite{a96} for more details).
Every $A_2$ singularity is locally of the form $\mathbb{C}^2/\Zbb_3$. It is given by modding out
$ (\xi,\eta) \sim (e^{2\pi i /3}\xi,e^{-2\pi i /3}\eta) $ from $\mathbb{C}^2$.
Let us choose three coordinates invariant under $\Zbb_3$:
\begin{equation}
 x_0 = \xi^3 \ ,\qquad x_1 = \eta^3 \ ,\qquad x_2 = \xi \eta\ ,
\end{equation}
which  obey 
\begin{equation}\label{EqA2Sing}
 x_0 x_1 - x_2^3 =0\ 
\end{equation}
in $\mathbb{C}^3$.
This space has one singularity located at the origin. One could remove this singularity by adding a term to the polynomial
(e.g.\ by deforming Eq.~\eqref{EqA2Sing} into $ x_0 x_1 - x_2^3 = \epsilon $). This would correspond 
to a complex structure deformation, i.e., to a rotation
of $\Omega_2$ into the $A_2$-cycles which then are of finite size. However, we are interested in a rotation of the K\"ahler form,
while keeping $\Omega_2$ orthogonal to the $A_2$-cycles. This
corresponds to blowing up the singularity. 
One introduces a $\mathbb{P}^2$ parametrized by coordinates $\{s_0,s_1,s_2\}$, subject to the conditions
\begin{equation}
 x_0 s_1 = x_1 s_0\ , \qquad x_0 s_2 = x_2 s_0 \ ,\qquad x_1 s_2 = x_2 s_1 \ .
\end{equation}
The $s_i$ are uniquely determined away from the origin $x_i=0$. On the other hand, at the point $x_i=0$ $\forall i$ they are left
undetermined. Let us choose the following path approaching the origin to find the exceptional divisor: $x_i= c_i t$, $t\rightarrow 0$.
To be inside the space, the following equation must hold:
\begin{equation}
 c_0 c_1 - t c_2^3 = 0 \qquad \Rightarrow \qquad s_0 s_1 = s_2^2 c_2
 t\ ,
\end{equation}
which for $t\rightarrow 0$ becomes
\begin{equation}
 s_0 s_1 = 0\ .
\end{equation}
We see that at $x_i=0$ there are two $\mathbb{P}^1$'s, $\{s_0=0\}$ and
$\{s_1=0\}$, intersecting at the point
$(x_0,x_1,x_2;$ $s_0,s_1,s_2)=(0,0,0;0,0,1)$.
These are the cycles with intersection matrix given in Eq.~\eqref{A2intesMat}. Moreover one can see that $\{s_0=0\}$ intersects the plane $\{\xi=0\}$
at the point $(x_0,x_1,x_2;s_0,s_1,s_2)=(0,0,0;0,1,0)$ and does not intersect $\{\eta=0\}$, while $\{s_1=0\}$ intersects the plane $\{\eta=0\}$
at the point $(x_0,x_1,x_2;s_0,s_1,s_2)=(0,0,0;1,0,0)$ and does not intersect $\{\xi=0\}$.

To summarize, we have an intuitive picture of the  smooth K3 which
arises from blowing up the orbifold $T^4/\Zbb_3$. This orbifold is an elliptic fibration over a $\mathbb{P}^1$ with nine $A_2$ singularities. Over a
generic point on the base, the fiber is a $T^2$. On top of three points we have the
singular fiber $T^2/\Zbb_3$. After the blow-up, the singular fiber becomes a
collection of seven (six independent) $\mathbb{P}^1$'s with intersection matrix given by the extended Dynkin diagram of $E_6$.
At this point of the moduli space we have a large number of algebraic cycles: the holomorphic two-form $\Omega_2$ is still
orthogonal to the full lattice
\begin{equation}\label{PXlattice2}
 P_X = {\cal U} \oplus (-E_6) \oplus (-E_6) \oplus (-E_6)\:.
\end{equation}
A generic smooth point in the K3 moduli space will have a smaller
Picard group. For the case at hand we allow $\Omega_2$ to move away 
from this symmetric point.

\subsection{Involution on the resolution of $T^4/\Zbb_3$}\label{AK3Invol}

As we have just seen, an involution on K3 is known once the lattice of even cycles is known. In the following we want to determine how the $\Zbb_2$ involution
on $T^4/\Zbb_3$ is promoted to a non-symplectic involution on the
smooth K3. To do so we  use the known action of the $\Zbb_2$ on the orbifold
and from this derive the transformation properties of the K3 cycles. In this
way we are able to determine the even lattice $S_+$ and its properties.

For this purpose, we will view $T^4/\Zbb_3$ as a singular K3 that is elliptically fibered over the base $T^2/\Zbb_3$. Over the singular point of the base (spanned by $z_1$) the fiber (spanned by $z_2$) degenerates to $T^2/\Zbb_3$. Let us call $z_1^a,z_1^b,z_1^c$ the points where the singular fiber sits. The two two-tori $\Pi_1$ and $\Pi_2$ are described in the following way: the first is just a point in the base of the elliptic fibration (i.e.\ it is homologous to the $T^2$ fiber), while the second is given by three points in the fiber, exchanged by $\Zbb_3$ (it wraps three times the base). After smoothing the singularities like above, the 18 exceptional $\mathbb{P}^1$'s $e^{(0)}_{\alpha\beta},e^{(1)}_{\alpha\beta}$ have finite size. The upper index is (1) if the exceptional cycle locally corresponds to $s_1=0$ and (0) if it corresponds to $s_0=0$.

Let us recall that the $\Zbb_2$ action on coordinates of $T^4$ is given by $(z_1,z_2)\mapsto (-z_1,z_2)$. Then the fiber of the elliptically fibered singular K3 is left invariant, while the $\Zbb_2$ group acts on the base and has two fixed points on it. 
Under $\Zbb_2$ the locations of the singular fibers behave as follows: $z_1^{\alpha=3} \leftrightarrow z_1^{\alpha=5}$ while $z_1^{\alpha=1}$ is left fixed (see Fig.~\ref{T4Z3fig}).
Let us call the other fixed point on the base $z_1^{\alpha=2,4,6}$ (on the $T^2$ spanned by $z_1$ it is given by three points that are exchanged under the $\Zbb_3$ transformation and that then are identified on the $T^2/\Zbb_3$ base).

From here we see immediately that the two-cycles corresponding to the fiber and the base of the fibration are even. In particular we have
a fixed two-torus over $z_1=z_1^{\alpha=2,4,6}$. Moreover, the exceptional cycles corresponding to $z_1=z_1^{\alpha=3}$ are exchanged with
the exceptional cycles corresponding to $z_1=z_1^{\alpha=5}$. To understand how the cycles on $z_1^c$ behaves
we have to use their algebraic description after the blow-up.
These cycles are given by the large singular fiber and the six exceptional cycles coming from blowing up the three $\mathbb{C}^2/\Zbb_3$ singularities on $z_1=z_1^{\alpha=1}$. The singular fiber is given by the equation $z_1=z_1^{\alpha=1}$ and is left fixed by $\Zbb_2$. Let us see the other cycles:
In the local coordinates around $z_1=z_1^c$, the action of $\Zbb_2$ is given by $(\xi,\eta)\mapsto (-\xi,\eta)$,
where $\xi\leftrightarrow z_1$ and $\eta\leftrightarrow z_2$. This implies the following
transformation on the $x_i$ and $s_i$:
\begin{equation}
 (x_0,x_1,x_2; s_0,s_1,s_2) \mapsto (-x_0,x_1,-x_2; s_0,-s_1,s_2)\ .
\end{equation}
We recall that the two $\mathbb{P}^1$'s at $x_i=0$ are determined by $s_0=0$ and $s_1=0$. From the transformation above, we see that the first
one transforms as the base of the elliptic fibration, i.e., it is even, with two fixed points, while the second is left fixed by $\Zbb_2$.
One of the fixed points of $\{s_0=0\}$ is the intersection point with $\{s_1=0\}$, while the other is the intersection point with the
singular fiber ($\xi=0$). As we have seen above, the base ($\eta=0$) intersects $\{s_1=0\}$, but neither the singular fiber nor $\{s_0=0\}$.

From these considerations we can determine both the even cycles and the fixed point locus. The even cycles are linear combination of:
\begin{itemize}
 \item the base and the fiber,
 \item the six exceptional cycles on $z_1^{\alpha=1}$,
 \item the six even combination of the exceptional cycles on $z_1^{\alpha=3}$ and $z_1^{\alpha=5}$.
\end{itemize}
As illustrated in Fig.~\ref{T4Z3fig}, the fixed locus is given by
\begin{itemize}
 \item the two-torus at $z_1=z_1^{\alpha=2,4,6}$,
 \item the three $\mathbb{P}^1$s $e^{(1)}_{1\beta}$ at $z_1=z_1^c$ and locally $s=1$.
\end{itemize}

\newpage

\section{4D Orbifold Spectrum after Higgsing}\label{HiggsedOrbifS}

Here we list the spectrum given in Table~\ref{OrbifSpectrum}, after the gauge group is broken to $SU(2)\times SU(3) \times SO(12)$. 
In the following table we also list the charges of the states with respect to the broken $U(1)$ generators given in \eqref{BrokenU1} and \eqref{BrokenU1bis}.

\begin{table}[h]
\label{OrbifSpectrumH0}
{\small
$$
\begin{array}{|lc|c|c|c|c|c|c|c|c|c|c|c|}
\hline
\mbox{Multiplet} && k & \alpha & \beta & \gamma & q_1^{(1)} & q_1^{(2)} & q_1^{(3)} & q_2 & q_3 & \tilde{q}_2 & \tilde{q}_4 \\
 \hline\hline
(1,1;32) && 0 & \mbox{---------} & --- & ---- & 0 & 0 & 0 & 0 & 0 & 1 & 2 \\
(1,1;32') && 0 & \mbox{---------} & --- & ---- & 0 & 0 & 0 & 0 & 0 & 1 & -1 \\
(2,3;1) && 0 & \mbox{---------} & --- & ---- & 3 & 3 & -3 & 0 & 1 & 0 & 0\\
(2,3;1) && 0 & \mbox{---------} & --- & ---- & 3 & 3 & -3 & 0 & -1 & 0 & 0\\
(2,\bar{3};1) && 0 & \mbox{---------} & --- & ---- & -3 & -3 & 3 & 0 & 1 & 0 & 0\\
(2,\bar{3};1) && 0 & \mbox{---------} & --- & ---- & -3 & -3 & 3 & 0 & -1 & 0 & 0\\
(2,1;1) && 0 & \mbox{---------} & --- & ---- & -3 & 3 & 3 & 0 & 1 & 0 & 0\\
(2,1;1) && 0 & \mbox{---------} & --- & ---- & -3 & 3 & -3 & 0 & 1 & 0 & 0\\
(2,1;1) && 0 & \mbox{---------} & --- & ---- & 3 & -3 & 3 & 0 & 1 & 0 & 0\\
(2,1;1) && 0 & \mbox{---------} & --- & ---- & 3 & -3 & -3 & 0 & 1 & 0 & 0\\
(2,1;1) && 0 & \mbox{---------} & --- & ---- & -3 & 3 & 3 & 0 & -1 & 0 & 0\\
(2,1;1) && 0 & \mbox{---------} & --- & ---- & -3 & 3 & -3 & 0 & -1 & 0 & 0\\
(2,1;1) && 0 & \mbox{---------} & --- & ---- & 3 & -3 & 3 & 0 & -1 & 0 & 0\\
(2,1;1) && 0 & \mbox{---------} & --- & ---- & 3 & -3 & -3 & 0 & -1 & 0 & 0\\
(1,3;1) && 0 & \mbox{---------} & --- & ---- & -5 & 1 & 1 & 1 & 1 & 0 & 0\\
(1,3;1) && 0 & \mbox{---------} & --- & ---- & 1 & -5 & 1 & 1 & 1 & 0 & 0\\
(1,3;1) && 0 & \mbox{---------} & --- & ---- & -5 & 1 & 1 & 1 & -1 & 0 & 0\\
(1,3;1) && 0 & \mbox{---------} & --- & ---- & 1 & -5 & 1 & 1 & -1 & 0 & 0\\
(1,\bar{3};1) && 0 & \mbox{---------} & --- & ---- & 1 & 1 & 1 & 1 & 1 & 0 & 0\\
(1,\bar{3};1) && 0 & \mbox{---------} & --- & ---- & 1 & 1 & -5 & 1 & 1 & 0 & 0\\
(1,\bar{3};1) && 0 & \mbox{---------} & --- & ---- & 1 & 1 & 1 & 1 & -1 & 0 & 0\\
(1,\bar{3};1) && 0 & \mbox{---------} & --- & ---- & 1 & 1 & -5 & 1 & -1 & 0 & 0\\
(1,1;1) && 0 & \mbox{---------} & --- & ---- & 1 & 7 & -5 & 1 & 1 & 0 & 0\\
(1,1;1) && 0 & \mbox{---------} & --- & ---- & -5 & -5 & 7 & 1 & 1 & 0 & 0\\
(1,1;1) && 0 & \mbox{---------} & --- & ---- & -5 & -5 & 1 & 1 & 1 & 0 & 0\\
(1,1;1) && 0 & \mbox{---------} & --- & ---- & 7 & 1 & -5 & 1 & 1 & 0 & 0\\
(1,1;1) && 0 & \mbox{---------} & --- & ---- & 1 & 7 & -5 & 1 & -1 & 0 & 0\\
(1,1;1) && 0 & \mbox{---------} & --- & ---- & -5 & -5 & 7 & 1 & -1 & 0 & 0\\
(1,1;1) && 0 & \mbox{---------} & --- & ---- & -5 & -5 & 1 & 1 & -1 & 0 & 0\\
(1,1;1) && 0 & \mbox{---------} & --- & ---- & 7 & 1 & -5 & 1 & -1 & 0 & 0\\
\hline 
\end{array}
$$}
\caption{\it Orbifold massless spectrum after Higgsing. Untwisted sector 
$k=0$.}
\end{table}
%
\vspace*{-1cm}
\begin{table}
\label{OrbifSpectrumH1}
{\small
$$
\begin{array}{|l|c|c|c|c|c|c|c|c|c|c|c|}
\hline
\mbox{Multiplet} & k & \alpha & \beta & \gamma & q_1^{(1)} & q_1^{(2)} & q_1^{(3)} & q_2 & q_3 & \tilde{q}_2 & \tilde{q}_4 \\
 \hline\hline
(2,3;1) & 0 & \mbox{---------} & --- & ---- & -4 & 2 & 2 & -1 & 0 & 0 & 0\\
(2,3;1) & 0 & \mbox{---------} & --- & ---- & 2 & -4 & 2 & -1 & 0 & 0 & 0\\
(2,\bar{3};1) & 0 & \mbox{---------} & --- & ---- & 2 & 2 & 2 & -1 & 0 & 0 & 0\\
(2,\bar{3};1) & 0 & \mbox{---------} & --- & ---- & 2 & 2 & -4 & -1 & 0 & 0 & 0\\
(2,1;1) & 0 & \mbox{---------} & --- & ---- & 2 & 8 & -4 & -1 & 0 & 0 & 0\\
(2,1;1) & 0 & \mbox{---------} & --- & ---- & -4 & -4 & 8 & -1 & 0 & 0 & 0\\
(2,1;1) & 0 & \mbox{---------} & --- & ---- & -4 & -4 & 2 & -1 & 0 & 0 & 0\\
(2,1;1) & 0 & \mbox{---------} & --- & ---- & 8 & 2 & -4 & -1 & 0 & 0 & 0\\
(1,1;12) & 0 & \mbox{---------} & --- & ---- & 0 & 0 & 0 & 0 & 0 & 2 & 1\\
(1,1;1) & 0 & \mbox{---------} & --- & ---- & 0 & 0 & 0 & 0 & 0 & 2 & 4\\
(1,1;1) & 0 & \mbox{---------} & --- & ---- & 0 & 0 & 0 & 0 & 0 & 2 & -2\\
(1,3;1) & 0 & \mbox{---------} & --- & ---- & 2 & 2 & -4 & 2 & 0 & 0 & 0\\
(1,\bar{3};1) & 0 & \mbox{---------} & --- & ---- & -4 & -4 & 2 & 2 & 0 & 0 & 0\\
(1,1;1) & 0 & \mbox{---------} & --- & ---- & -4 & 2 & 2 & 2 & 0 & 0 & 0\\
(1,1;1) & 0 & \mbox{---------} & --- & ---- & -4 & 2 & -4 & 2 & 0 & 0 & 0\\
(1,1;1) & 0 & \mbox{---------} & --- & ---- & 2 & -4 & 2 & 2 & 0 & 0 & 0\\
(1,1;1) & 0 & \mbox{---------} & --- & ---- & 2 & -4 & -4 & 2 & 0 & 0 & 0\\
(2,1;1) & 0 & \mbox{---------} & --- & ---- & 1 & 1 & 1 & -2 & 1 & 0 & 0\\
(2,1;1) & 0 & \mbox{---------} & --- & ---- & 1 & 1 & 1 & -2 & -1 & 0 & 0\\
 \hline\hline
(1,3;1) & 1 & 1 & 1,2,3 & 1,2,3,4 & -13/3 & 5/3 & 5/3 & -1/3 & 0 & 2/3 & 1/3\\ 
(1,3;1) & 1 & 1 & 1,2,3 & 1,2,3,4 &  5/3 & -13/3 & 5/3 & -1/3 & 0 & 2/3 & 1/3\\
(1,\bar{3};1) & 1 & 1 & 1,2,3 & 1,2,3,4 &  5/3 & 5/3 & 5/3 & -1/3 & 0 & 2/3 & 1/3\\
(1,\bar{3};1) & 1 & 1 & 1,2,3 & 1,2,3,4 & 5/3 & 5/3 & -13/3 & -1/3 & 0 & 2/3 & 1/3\\
(1,1;1) & 1 & 1 & 1,2,3 & 1,2,3,4 & 5/3 & 23/3 & -13/3 & -1/3 & 0 & 2/3 & 1/3\\
(1,1;1) & 1 & 1 & 1,2,3 & 1,2,3,4 & -13/3 & -13/3 & 23/3 & -1/3 & 0 & 2/3 & 1/3\\
(1,1;1) & 1 & 1 & 1,2,3 & 1,2,3,4 & -13/3 & -13/3 & 5/3 & -1/3 & 0 & 2/3 & 1/3\\
(1,1;1) & 1 & 1 & 1,2,3 & 1,2,3,4 & 23/3 & 5/3 & -13/3 & -1/3 & 0 & 2/3 & 1/3\\
(2,1;1) & 1 & 1 & 1,2,3 & 1,2,3,4 & -1/3 & -1/3 & -1/3 & 2/3 & 0 & 2/3 & 1/3\\
(2,1;1) & 1 & 1 & 1,2,3 & 1,2,3,4 & -1/3 & -1/3 & -1/3 & 2/3 & 0 & 2/3 & 1/3\\
(1,1;1) & 1 & 1 & 1,2,3 & 1,2,3,4 & 2/3 & 2/3 & 2/3 & -4/3 & 1 & 2/3 & 1/3\\
(1,1;1) & 1 & 1 & 1,2,3 & 1,2,3,4 & 2/3 & 2/3 & 2/3 & -4/3 & -1 & 2/3 & 1/3\\
\hline \hline
(1,1;1) & 2 & 1 & 1,2,3 & - & -2/3 & -2/3 & -2/3 & 4/3 & 0 & 4/3 & 2/3 \\
(2,1;1) & 2 & 1 & 1,2,3 & - & 1/3 & 1/3 & 1/3 & -2/3 & 1 & 4/3 & 2/3 \\
(2,1;1) & 2 & 1 & 1,2,3 & - & 1/3 & 1/3 & 1/3 & -2/3 & -1 & 4/3 & 2/3 \\
\hline
(1,1;1) & 2 & 3+5 & 1,2,3 & - & -2/3 & -2/3 & -2/3 & 4/3 & 0 & 4/3 & 2/3 \\
(2,1;1) & 2 & 3+5 & 1,2,3 & - & 1/3 & 1/3 & 1/3 & -2/3 & 1 & 4/3 & 2/3 \\
(2,1;1) & 2 & 3+5 & 1,2,3 & - & 1/3 & 1/3 & 1/3 & -2/3 & -1 & 4/3 & 2/3 \\
\hline
(1,1;1) & 2 & 3-5 & 1,2,3 & - & -2/3 & -2/3 & -2/3 & 4/3 & 0 & 4/3 & 2/3 \\
(1,3;1) & 2 & 3-5 & 1,2,3 & - & 10/3 & 10/3 & -8/3 & -2/3 & 0 & 4/3 & 2/3 \\
(1,\bar{3};1) & 2 & 3-5 & 1,2,3 & - & -8/3 & -8/3 & 10/3 & -2/3 & 0 & 4/3 & 2/3 \\
(1,1;1) & 2 & 3-5 & 1,2,3 & - & -8/3 & 10/3 & 10/3 & -2/3 & 0 & 4/3 & 2/3 \\
(1,1;1) & 2 & 3-5 & 1,2,3 & - & -8/3 & 10/3 & -8/3 & -2/3 & 0 & 4/3 & 2/3 \\
\hline
\end{array}
$$}
\vspace*{-5mm}
\caption{\it Orbifold massless spectrum after higgsing. Untwisted sector
$k=0$ continued; twisted sectors $k=1$ and $k=2$.}
\end{table}

\begin{table}{}
\label{OrbifSpectrumH2}
{\small
$$
\begin{array}{|l|c|c|c|c|c|c|c|c|c|c|c|}
\hline
\mbox{Multiplet} & k & \alpha & \beta & \gamma & q_1^{(1)} & q_1^{(2)} & q_1^{(3)} & q_2 & q_3 & \tilde{q}_2 & \tilde{q}_4 \\
\hline \hline
(1,1;1) & 2 & 3-5 & 1,2,3 & - & 10/3 & -8/3 & 10/3 & -2/3 & 0 & 4/3 & 2/3 \\
(1,1;1) & 2 & 3-5 & 1,2,3 & - & 10/3 & -8/3 & -8/3 & -2/3 & 0 & 4/3 & 2/3 \\
(1,1;12) & 2 & 3-5 & 1,2,3 & - & -2/3 & -2/3 & -2/3 & 4/3 & 0 & -2/3 & -1/3 \\
(1,1;1) & 2 & 3-5 & 1,2,3 & - & -2/3 & -2/3 & -2/3 & 4/3 & 0 & -2/3 & 8/3 \\
(1,1;1) & 2 & 3-5 & 1,2,3 & - & -2/3 & -2/3 & -2/3 & 4/3 & 0 & -2/3 & -10/3 \\
\hline\hline
(1,1;1) & 3 & 1 & - & 1,2,3,4 & 0 & 0 & 0 & 0 & 1 & 2 & 1  \\
(1,1;1) & 3 & 1 & - & 1,2,3,4 & 0 & 0 & 0 & 0 & -1 & 2 & 1  \\
\hline
(1,1;1) & 3 & 2+4+6 & - & 1,2,3,4 & 0 & 0 & 0 & 0 & 1 & 2 & 1 \\
(1,1;1) & 3 & 2+4+6 & - & 1,2,3,4 & 0 & 0 & 0 & 0 & -1 & 2 & 1 \\
\hline
(1,1;1) & 3 & 2+\nu 4+\nu^2 6 & - & 1,2,3,4 & 0 & 0 & 0 & 0 & 1 & -2 & -1 \\
(1,1;1) & 3 & 2+\nu 4+\nu^2 6 & - & 1,2,3,4 & 0 & 0 & 0 & 0 & -1 & -2 & -1 \\
\hline
(1,1;12) & 3 & 2+\nu^2 4+\nu^4 6 & - & 1,2,3,4 & 0 & 0 & 0 & 0 & 1 & 0 & 0 \\
(1,1;12) & 3 & 2+\nu^2 4+\nu^4 6 & - & 1,2,3,4 & 0 & 0 & 0 & 0 & -1 & 0 & 0 \\
(1,1;1) & 3 & 2+\nu^2 4+\nu^4 6 & - & 1,2,3,4 & 0 & 0 & 0 & 0 & 1 & 0 & 3 \\
(1,1;1) & 3 & 2+\nu^2 4+\nu^4 6 & - & 1,2,3,4 & 0 & 0 & 0 & 0 & -1 & 0 & 3 \\
(1,1;1) & 3 & 2+\nu^2 4+\nu^4 6 & - & 1,2,3,4 & 0 & 0 & 0 & 0 & 1 & 0 & -3 \\
(1,1;1) & 3 & 2+\nu^2 4+\nu^4 6 & - & 1,2,3,4 & 0 & 0 & 0 & 0 & -1 & 0 & -3 \\
 \hline\hline
(1,1;1) & 4 & 1 & 1,2,3 & - & 2/3 & 2/3 & 2/3 & -4/3 & 0 & -4/3 & -2/3 \\
(1,3;1) & 4 & 1 & 1,2,3 & - & 8/3 & 8/3 & -10/3 & 2/3 & 0 & -4/3 & -2/3 \\
(1,\bar{3};1) & 4 & 1 & 1,2,3 & - & -10/3 & -10/3 & 8/3 & 2/3 & 0 & -4/3 & -2/3 \\
(1,1;1) & 4 & 1 & 1,2,3 & - & -10/3 & 8/3 & 8/3 & 2/3 & 0 & -4/3 & -2/3 \\
(1,1;1) & 4 & 1 & 1,2,3 & - & -10/3 & 8/3 & -10/3 & 2/3 & 0 & -4/3 & -2/3 \\
(1,1;1) & 4 & 1 & 1,2,3 & - & 8/3 & -10/3 & 8/3 & 2/3 & 0 & -4/3 & -2/3 \\
(1,1;1) & 4 & 1 & 1,2,3 & - & 8/3 & -10/3 & -10/3 & 2/3 & 0 & -4/3 & -2/3 \\
(1,1;12) & 4 & 1 & 1,2,3 & - & 2/3 & 2/3 & 2/3 & -4/3 & 0 & 2/3 & 1/3 \\
(1,1;1) & 4 & 1 & 1,2,3 & - & 2/3 & 2/3 & 2/3 & -4/3 & 0 & 2/3 & 10/3 \\
(1,1;1) & 4 & 1 & 1,2,3 & - & 2/3 & 2/3 & 2/3 & -4/3 & 0 & 2/3 & -8/3 \\
\hline
(1,1;1) & 4 & 3+5 & 1,2,3 & - & 2/3 & 2/3 & 2/3 & -4/3 & 0 & -4/3 & -2/3 \\
(1,3;1) & 4 & 3+5 & 1,2,3 & - & 8/3 & 8/3 & -10/3 & 2/3 & 0 & -4/3 & -2/3 \\
(1,\bar{3};1) & 4 & 3+5 & 1,2,3 & - & -10/3 & -10/3 & 8/3 & 2/3 & 0 & -4/3 & -2/3 \\
(1,1;1) & 4 & 3+5 & 1,2,3 & - & -10/3 & 8/3 & 8/3 & 2/3 & 0 & -4/3 & -2/3 \\
(1,1;1) & 4 & 3+5 & 1,2,3 & - & -10/3 & 8/3 & -10/3 & 2/3 & 0 & -4/3 & -2/3 \\
(1,1;1) & 4 & 3+5 & 1,2,3 & - & 8/3 & -10/3 & 8/3 & 2/3 & 0 & -4/3 & -2/3 \\
(1,1;1) & 4 & 3+5 & 1,2,3 & - & 8/3 & -10/3 & -10/3 & 2/3 & 0 & -4/3 & -2/3 \\
(1,1;12) & 4 & 3+5 & 1,2,3 & - & 2/3 & 2/3 & 2/3 & -4/3 & 0 & 2/3 & 1/3 \\
(1,1;1) & 4 & 3+5 & 1,2,3 & - & 2/3 & 2/3 & 2/3 & -4/3 & 0 & 2/3 & 10/3 \\
(1,1;1) & 4 & 3+5 & 1,2,3 & - & 2/3 & 2/3 & 2/3 & -4/3 & 0 & 2/3 & -8/3 \\
\hline
(1,1;1) & 4 & 3-5 & 1,2,3 & - & 2/3 & 2/3 & 2/3 & -4/3 & 0 & -4/3 & -2/3 \\
(2,1;1) & 4 & 3-5 & 1,2,3 & - & -1/3 & -1/3 & -1/3 & 2/3 & 1 & -4/3 & -2/3 \\
(2,1;1) & 4 & 3-5 & 1,2,3 & - & -1/3 & -1/3 & -1/3 & 2/3 & -1 & -4/3 & -2/3 \\
\hline\hline
\end{array}
$$}
\caption{\it Orbifold massless spectrum after Higgsing. Twisted sectors
$k=2$ (continued), $k=3$ and $k=4$. }
\end{table}

\newpage

\section{4D Spectrum in the Smooth Case}\label{chirals}

In this appendix we compute the spectrum of 
the heterotic $E_8\times E_8$ string compactified on the Voisin-Borcea manifold $Y$ which we discussed in Section~\ref{smoothS} together with the choice 
of the flux given in Eqs.~\eqref{flux} and \eqref{fluxsol}.
The unbroken gauge group is 
\begin{equation}
SU(2)\times SU(3) \times U(1)^5 \times SO(12)\times U(1)^2 \ ,
\end{equation}
where the generators of the five $U(1)$'s in the first $E_8$ are chosen to be
\begin{align}\label{BrokenU1}
&T_1^{(1)} = (0,0,-1,-3,-3,-3,-3,-3)\ ,  &\qquad&  T_1^{(2)} = (0,0,-1,+3,+3,-3,-3,-3)\ ,  \nonumber\\
&T_1^{(3)} = (0,0,-1,+3,-3,+3,+3,+3)\ ,  &\qquad&  T_2 = (0,0,2,0,0,0,0,0)\ ,  \\
 &T_3 = (+1,-1,0,0,0,0,0,0)\ ,  &\qquad&   \nonumber
\end{align}
and the generators of the two $U(1)$'s in the second $E_8$ are
\begin{align}\label{BrokenU1bis}
&\tilde{T}_2 = (2,0,0,0,0,0,0,0) \ , &\qquad&  \tilde{T}_4 = (1,3,0,0,0,0,0,0)\ .  
\end{align}
\begin{table}
$$
\begin{array}{|l|c|c|c|c|c|c||l|c|c|c|c|c|c|}
\hline
 {\rm State} & q_1^{(1)} & q_1^{(2)} & q_1^{(3)} & q_2 & q_3 & n & {\rm State} & q_1^{(1)} & q_1^{(2)} & q_1^{(3)} & q_2 & q_3 & n \\ \hline 
 {\bf (3,1)} & -5 & 1 & 1 & 1 & 1 & -1 & {\bf (3,2)} & -2 & -2 & -2 & 1 & 0 & 1 \\
 {\bf (3,1)} & 1 & -5 & 1 & 1 & 1 & -1 &  {\bf (3,2)} & -2 & -2 & 4 & 1 & 0 & 1 \\
 {\bf (\bar{3},1)} & 1 & 1 & -5 & 1 & 1 & -1 & {\bf (\bar{3},2)} & -2 & 4 & -2 & 1 & 0 & 1 \\
 {\bf (\bar{3},1)} & 1 & 1 & 1 & 1 & 1 & -1 & {\bf (\bar{3},2)} & 4 & -2 & -2 & 1 & 0 & 1 \\
 {\bf (1,1)} & -5 & -5 & 7 & 1 & 1 & 5 & {\bf (1,2)} & -8 & -2 & 4 & 1 & 0 & -5 \\
 {\bf (1,1)} & -5 & -5 & 7 & 1 & 1 & 5 & {\bf (1,2)} & -2 & -8 & 4 & 1 & 0 & -5 \\
 {\bf (1,1)} & 7 & 1 & -5 & 1 & 1 & 5 & {\bf (1,2)} & 4 & 4 & -8 & 1 & 0 & -5 \\
 {\bf (1,1)} & 1 & 7 & -5 & 1 & 1 & 5 & {\bf (1,2)} & 4 & 4 & -2 & 1 & 0 & -5 \\
 {\bf (3,1)} & -5 & 1 & 1 & 1 & -1 & -1 & {\bf (1,2)} & -1 & -1 & -1 & 2 & 1 & 4 \\
 {\bf (3,1)} & 1 & -5 & 1 & 1 & -1 & -1 & {\bf (1,2)} & -1 & -1 & -1 & 2 & -1 & 4 \\
 {\bf (\bar{3},1)} & 1 & 1 & -5 & 1 & -1 & -1 & {\bf (3,1)} & 2 & 2 & -4 & 2 & 0 & 2 \\
 {\bf (\bar{3},1)} & 1 & 1 & 1 & 1 & -1 & -1 & {\bf (\bar{3},1)} & -4 & -4 & 2 & 2 & 0 & 2 \\
 {\bf (1,1)} & -5 & -5 & 7 & 1 & -1 & 5 & {\bf (1,1)} & 2 & -4 & -4 & 2 & 0 & 2 \\
 {\bf (1,1)} & -5 & -5 & 7 & 1 & -1 & 5 & {\bf (1,1)} & -4 & 2 & -4 & 2 & 0 & 2 \\
 {\bf (1,1)} & 7 & 1 & -5 & 1 & -1 & 5 & {\bf (1,1)} & -4 & 2 & 2 & 2 & 0 & 2 \\
 {\bf (1,1)} & 1 & 7 & -5 & 1 & -1 & 5 & {\bf (1,1)} & 2 & -4 & 2 & 2 & 0 & 2 \\ 
 {\bf (3,1)} & 0 & 6 & -6 & 0 & 0 & 0  & {\bf (1,1)} & 0 & 0 & 0 & 0 & 2 & 4 \\
 {\bf (3,1)} & 0 & 6 & 0 & 0 & 0 & 0 & {\bf (3,2)} & 3 & 3 & -3 & 0 & -1 & 0 \\
 {\bf (\bar{3},1)} & -6 & 0 & 6 & 0 & 0 & 0 & {\bf (\bar{3},2)} & -3 & -3 & 3 & 0 & -1 & 0 \\
 {\bf (\bar{3},1)} & -6 & 0 & 0 & 0 & 0 & 0 & {\bf (1,2)} & 3 & -3 & -3 & 0 & -1 & 0 \\
 {\bf (1,1)} & -6 & 6 & 0 & 0 & 0 & 0 & {\bf (1,2)} & 3 & -3 & 3 & 0 & -1 & 0 \\
 {\bf (1,1)} & 0 & 0 & 6 & 0 & 0 & 0 & {\bf (1,2)} & -3 & 3 & 3 & 0 & -1 & 0 \\
 {\bf (3,1)} & -6 & -6 & 6 & 0 & 0 & 0 & {\bf (1,2)} & -3 & 3 & -3 & 0 & -1 & 0 \\\hline
\end{array}
$$
\caption{\it Representation ${\bf R}^{(1)}$ in the decomposition \eqref{1stE8Adj} of the first $E_8$.}\label{firstE8}
\end{table}

We now list the spectrum which arises from the decomposition 
of the adjoint representation of the two $E_8$ factors,
\begin{eqnarray}
\label{1stE8Adj} {\bf 248}^{(1)} &=& ({\bf 1},{\bf 3})\oplus ({\bf 8},{\bf 1}) 
\oplus ({\bf 1},{\bf 1})^5 
 \oplus ({\bf R^{(1)}}\oplus {\bf \bar{R}^{(1)}})\ ,\\
\label{2ndE8Adj} {\bf 248}^{(2)} &=& {\bf 66}  \oplus {\bf 1} \oplus {\bf 1} \oplus ({\bf R^{(2)}}\oplus {\bf \bar{R}^{(2)}}) \ .
\end{eqnarray}
Each state is given together with its charges under
the $U(1)$ generators \eqref{BrokenU1} and \eqref{BrokenU1bis}. Moreover, the chiral index is given for each state. It
is obtained by applying the number operator \eqref{Nop} to the state.
Since the fluxes $f_I$ are linear combinations of the above $U(1)$ generators, the chiral index will be give in terms of the charges. For the case under study,
the chiral multiplicity of a state of charges $(q_1^{(1)},q_1^{(2)},q_1^{(3)},q_2,q_3,\tilde{q}_2,\tilde{q}_4)$ is
 \begin{equation}\begin{aligned}
   n =&  \tfrac{1}{108} \left[52 q_2^3 + 
   3 q_2 ((q_1^{(1)})^2 + (q_1^{(2)})^2 + (q_1^{(3)})^2 + (q_1^{(1)} + q_1^{(2)} + q_1^{(3)} + 2 q_2)^2) \right. \\ 
&+ 
   2 ((q_1^{(1)})^3 + (q_1^{(2)})^3 + (q_1^{(3)})^3 - (q_1^{(1)} + q_1^{(2)} + q_1^{(3)} + 2 q_2)^3) + 
   108 q_2 q_3^2  \\ 
& \left.  + 72 q_3^3 - 36 (3 q_2 + 2 q_3)\right]
   -\tfrac13 \left[5 \tilde{q}_2 + 2 \tilde{q}_2^3 + 3\tilde{q}_4 - 3\tilde{q}_2^2 \tilde{q}_4 \right]\ .
 \end{aligned}\end{equation}

\begin{table}
$$
\begin{array}{|r|c|c|c||r|c|c|c||r|c|c|c|}
\hline
 {\rm State} & \tilde{q}_2 & \tilde{q}_4 & n & {\rm State} & \tilde{q}_2 & \tilde{q}_4 & n & {\rm State} & \tilde{q}_2 & \tilde{q}_4 & n \\ \hline 
 {\bf 32} & 1 & 2 & -1 & {\bf 12} & 2 & 1 & -1 & {\bf 1} & 2 & 4 & -10 \\
 {\bf 32'} & 1 & -1 & -1 & {\bf 12} & 0 & 3 & 3 & {\bf 1} & 2 & -2 & 8 \\
\hline\end{array}
$$
\caption{\it Representation ${\bf R}^{(2)}$ in the decomposition \eqref{2ndE8Adj} of the first $E_8$.}\label{secondE8}
\end{table}

In Tables~\ref{firstE8} (\ref{secondE8}) we display the states which 
arise from the decomposition of the adjoint 
representation of the first (second) $E_8$. 
For the first $E_8$ all of them have $\tilde{q}_2=0$ and $\tilde{q}_4=0$,
while for the second $E_8$
all of them have $q_1^{i}=0$ and $q_2=q_3=0$.
The adjoint decomposes into the sum of the adjoint representations 
of the unbroken group, plus the sum of the reducible representation ${\bf R}$ 
and its complex conjugate ${\bf \bar{R}}$. In both tables
we only list the irreducible representation coming from ${\bf R}$.

\end{document}